\documentclass[11pt,a4paper]{article}

\usepackage[T1]{fontenc}
\usepackage[utf8]{inputenc}
\usepackage{lmodern}

\usepackage{amsmath}
\usepackage{amssymb}
\usepackage{dsfont}

\usepackage{natbib}

\usepackage{subcaption}

\usepackage{orcidlink}
\usepackage[normalem]{ulem}
\usepackage{framed}
\usepackage{tikz}

\usetikzlibrary{arrows.meta}

\usepackage[margin=2.5cm]{geometry}

\usepackage{setspace}
\usepackage{hyperref}
\hypersetup{hidelinks}

\newcommand{\proglang}[1]{\textsf{#1}}
\newcommand{\pkg}[1]{\texttt{#1}}
\newcommand{\code}[1]{\texttt{#1}}
\newcommand{\fct}[1]{\texttt{#1()}}

\usepackage{tcolorbox}
\tcbuselibrary{listings,breakable}
\usepackage{listings}
\usepackage{xcolor}

\definecolor{codebg}{RGB}{248,248,248}
\definecolor{codecomment}{RGB}{100,100,100}
\definecolor{codekeyword}{RGB}{0,80,160}
\definecolor{codestring}{RGB}{160,60,20}

\lstdefinelanguage{R}{
  keywords={
    if,else,for,while,function,repeat,in,next,break,
    TRUE,FALSE,NULL,NA,NaN,Inf
  },
  sensitive=true,
  comment=[l]{\#},
  morestring=[b]",
  morestring=[b]'
}

\newtcblisting{Code}{
  listing only,
  breakable,
  colback=codebg,
  colframe=black!20,
  listing options={
    language=R,
    basicstyle=\ttfamily\small,
    keywordstyle=\color{codekeyword},
    commentstyle=\color{codecomment},
    stringstyle=\color{codestring},
    breaklines=true,
    breakatwhitespace=false,
    columns=fullflexible,
    keepspaces=true,
    showstringspaces=false,
    frame=none
  },
  boxrule=0.5pt,
  arc=2pt,
  left=6pt,
  right=6pt,
  top=5pt,
  bottom=5pt
}

\title{The \proglang{R} package \pkg{LSJM} for fitting location-scale
joint models for a longitudinal marker and complex survival data}

\author{
Léonie Courcoul\\
Univ. Bordeaux\\
INSERM U1219\\
Bordeaux Population Health
\and
Hélène Jacqmin-Gadda\\
Univ. Bordeaux\\
INSERM U1219\\
Bordeaux Population Health
\and
Antoine Barbieri\\
Univ. Bordeaux\\
INSERM U1219\\
Bordeaux Population Health
}

\date{}

\begin{document}

\maketitle

\begin{abstract}

The \pkg{LSJM} \proglang{R} package provides a series of functions to
estimate models based on location-scale linear mixed model and location-scale
joint model theory. For longitudinal data, it includes the estimation of
location-scale mixed models with subject-specific residual variance that can
be time-dependent or distinguish within-visit and between-visits variability.
When survival data is considered, implemented joint models handle competing
risks as well as semi-competing risks possibly with interval-censored data
through the use of a illness-death sub-model. They incorporate flexible
dependence structure between the longitudinal marker and the times-to-events
through the current value, the slope and/or the residual variability of the
marker. Maximum likelihood estimators are obtained using the
Marquardt-Levenberg algorithm with stringent convergence criteria. The package
also provides several post-fit functions including goodness-of-fit analyses
and individual dynamic predictions. Some examples on real data allows to
illustrates the use of this package.

\end{abstract}


\proglang{R}

\vspace{1em}


%


\section*{Author information}

\noindent
\textbf{Léonie Courcoul}, \textbf{Hélène Jacqmin-Gadda},
\textbf{Antoine Barbieri}\\
Univ. Bordeaux, France\\
INSERM U1219\\
Bordeaux Population Health\\
E-mail:
\href{mailto:courcoul.leonie498@gmail.com}{courcoul.leonie498@gmail.com},
\href{mailto:helene.jacqmin-gadda@u-bordeaux.fr}{helene.jacqmin-gadda@u-bordeaux.fr},
\href{mailto:antoine.barbieri@u-bordeaux.fr}{antoine.barbieri@u-bordeaux.fr}

\section{Introduction}
Joint modeling of longitudinal and time-to-event data has become an essential tool to simultaneously analyze longitudinal data (measured repeatedly over time) and survival events (such as disease occurrence or death) \citep{wulfsohn_1997, tsiatis_2004}. The strength of joint models relies on their ability to capture the relationship between the evolution of a biomarker and the risk of an event. Traditional approaches that analyse these two types of data separately often miss critical interactions, whereas joint models integrate both, offering a more comprehensive understanding of how an individual's biomarker trajectory influences his risk of future events. The joint models combine a mixed model fitting the mean behavior over time of the longitudinal outcome and a survival model for the risk of developing one or more health events. Their estimations are based on the joint likelihood of these two models, whose dependency structure defines the type of joint model used \citep{wulfsohn_1997, proust_2009}. The first concerns latent class joint models \citep{proust_2009}. They assume that the population of interest is heterogeneous, and that there are homogeneous sub-populations that differ both on the mean trajectory of the longitudinal marker and on the risk of developing the event of interest. Conditionally on class membership, the longitudinal variable and event times are considered independent. The second type of joint models concerns shared-random effect models \citep{rizopoulos_2012}. Here, the dependence structure is defined by one or more functions of the random effects from the mixed model which are included as covariates in the survival model. They thus make it possible to assess the impact of a characteristic of the mean trajectory of the marker on the risk of developing the event. In this approach, the longitudinal variable and time-to-event are assumed to be independent conditionally on the random effects.\\
Over the years, several \proglang{R} packages have been developed to fit joint models for longitudinal and time-to-event data, each offering specific features and varying estimation methods. The \pkg{JM} package \citep{rizopoulos_2010} was one of the first \proglang{R} packages designed for fitting joint models with shared random effects. It enables the estimation of a joint model with a single event risk, using only the current value as the association function between the two submodels. The \pkg{JMbayes} package \citep{rizopoulos_2016} relies on Bayesian methods using Markov Chain Monte Carlo (MCMC) techniques to estimate the posterior distributions of model parameters. This package allows for the consideration of different types of longitudinal outcomes by using a generalized linear mixed model for the longitudinal part. The baseline risk is modeled using penalized B-splines, and left truncation, as well as time-dependent and exogenous covariates, can be accounted for in the survival submodel. Furthermore, various associations between the marker and survival can be specified, such as the current value, the slope, the cumulative effect or even the simple sharing of subject-specific random effects. The integral in the survival function is computed using either Gauss-Kronrod or Gauss-Legendre quadrature. The \pkg{JMbayes2} package \citep{rizopoulos_2023} is an evolution of \pkg{JMbayes}, offering a broader range of longitudinal and survival submodels. It allows for the estimation of multi-marker joint models and the inclusion of competing risks. In a frequentist framework, the \pkg{joineR} package \citep{philipson_2018} allows to fit joint model by maximizing the likelihood using EM algorithm and Gauss-Hermite integration on the random effects. The authors considered a semi-parametric approach, meaning that the baseline hazard is unspecified. It also handles competing risks, and a second version of the package, \pkg{joineRML} \citep{hickey_2018}, allows for multi-marker models, using Monte Carlo integration on the random effects. The \pkg{rstanarm} package \citep{brilleman_2018} also provides a Bayesian approach using advanced sampling techniques, such as the No-U-Turn Sampler (NUTS). The \pkg{INLAjoint} package \citep{rustand_2023} introduces an alternative Bayesian approach by leveraging Integrated Nested Laplace Approximation (INLA) for fast estimation. It supports multiple markers with different distributions, as well as competing risks and multi-state models. Lastly, the \pkg{lcmm} package \citep{proustlima_2017} allows the estimation of latent class joint models, including competing risks, by maximizing the likelihood using the Marquardt-Levenberg algorithm \citep{levenberg_1944, marquardt_1963}. Even though several software handle joint models for competing risks or multi-state data, none of them allows the estimation of joint models with interval-censored semi-competing events in shared-random effects model approach. Semi-competiting events occur when an event (as death) prevents the event of interest (e.g., a disease) from occurring, but the competing event can still occur after the event of interest (e.g., death after the disease). In the presence of semi-competing events, an illness-death model is then necessary. The estimation of such a model is complicated when the disease may be interval censored, for instance when the diagnosis may be assessed only at prespecified visits. Indeed, whether the disease and the death arise between two visits, the disease can not be diagnosed. This is a frequent problem in cohort studies about dementia in the elderly. \\
Another strong assumption used in all existing packages for shared random effect models is that the residual variance in mixed model is assumed homogeneous. This is not hold in more complex data settings. More recently, \cite{gao_2011} and \cite{barrett_2019} have extended the joint model with shared random effects to include a subject-specific residual variability of the marker in the risk of the event. These models combine a location-scale linear mixed model and a proportional hazard model. They have been motivated by the study of the impact of the within individual variability of blood pressure on the risk of cardiovascular events and by the study of the variability of intraocular pressure as an independent risk of primary open-angle glaucoma. In medical research, many current hypotheses regards the role of the variability of markers or risk factors (blood pressure, glycemia...) on the risk of clinical events (cardiovascular diseases, dementia...). However, the models proposed by \cite{gao_2011} and \cite{barrett_2019} have strong constraints (constant residual variance, no competing risks...) and no package was proposed for estimating these models.\\
On the other hand, several software exist to estimate linear location-scale mixed models that accounts for heteroscedasticity in the data by modeling the residual variance as a function of covariates or time. Softwares \pkg{MIXREGLS} \citep{hedecker_2013}, \pkg{MixWild} \citep{dzubur_2020} and the \proglang{R}-package \pkg{LMMEL} \citep{martin_2024} have been proposed to estimate such kind of models. However, they do not support joint modeling to study the impact of a marker's residual variability on event risk.\\
To address the need for more flexible joint models, we developed the \pkg{LSJM} package, designed to estimate joint models with subject-specific residual variance while addressing the challenges discussed in \cite{courcoul_SIM_2025} and \cite{courcoul2_2024}. Firstly, this package enables the modeling of continuous longitudinal data with heteroscedasticity, allowing both the mean trajectory and the residual variance, to depend on time, covariates and random effects. Secondly, it supports joint modeling of a continuous longitudinal marker with heteroscedasticity, combined with survival data. The survival component is highly flexible, allowing for the modeling of a single event risk, competing risks, or semi-competing risks, possibly interval censored, using cause-specific models (including illness-death models). As with residual variance modeling, the last survival feature is unique among packages estimating joint models with shared-random effects. Additionally, the joint model estimation accommodates left truncation and goodness-of-fit analyses and computation of individual risk in prediction are implemented in the package. Package \pkg{LSJM} is
available from the Comprehensive R Archive Network (CRAN) at \url{https://CRAN.R-project.org/package=LSJM}.\\
The paper is organized as follows. Section 2 defines the statistical models implemented in the package and Section 3 details the estimation process. Section 4 describes the post-fit analyses. Section 5 details the implementation of the estimation functions and the post-fit functions. Finally, Section 6 provides two examples based on the \code{threeC} dataset, a random subsample of 500 subjects from the French Three-Cities cohort \citep{3c_2003} available in the package and Section 7 concludes.

\section{Location-Scale Joint Models}
This section describes each family of statistical models implemented in the \pkg{LSJM} package. The first subsection focuses on the location-scale linear mixed models including the standard linear mixed model. It presents the general notations of the longitudinal model. The second subsection is dedicated to the presentation of the location-scale joint models for jointly analyzing a longitudinal marker with heteroscedasticity and complex survival data.  


\subsection{The location-scale mixed models (LSMM)}

\subsubsection{The standard linear mixed model}
\label{LMM}
Let us consider a sample of $N$ individuals. For each individual $i \in \{1,...,N\}$, we consider the $n_i$-vector of repeated measures $Y_i = (Y_{i1},...,Y_{in_i})^\top$ with $Y_{ij}$ the value of the longitudinal outcome of individual $i$ at time $t_{ij}$ $(j = 1,\ldots,n_i)$. \cite{laird_1982} proposed the standard linear mixed model:
\begin{equation}
    Y_{ij} = Y_{i}(t_{ij}) = \widetilde{Y}_i(t_{ij}) + \epsilon_{ij} = X_{ij}^{\top} \beta+Z_{ij}^{\top} b_{i}+\epsilon_{i}(t_{ij})
    \label{Mixed0}
\end{equation}
where $X_{ij}$ and $Z_{ij}$ are vectors of explanatory variables for subject $i$ at visit $j$, respectively associated with the fixed-effect vector $\beta$ and the subject-specific random-effect vector $b_i$ such as $b_i \sim \mathcal{N}(0,\Sigma_b)$, with $\Sigma_b$ an unspecified matrix. The residual errors $\epsilon_{ij}$ are assumed to be independent of $b_i$ and $\epsilon_{ij}\sim\mathcal{N}\left(0,\sigma^2_\epsilon\right)$. 

\subsubsection{LSMM with time-dependent variability}
Some authors have proposed to overcome the homoscedasticity assumption in the linear mixed model by defining the residual variance as a function of covariates and random effects \citep{aitkin_1987,foulley_1992,lin_1997,hedeker_2008}. These models are typically called location-scale mixed model (LSMM). For longitudinal measurements assuming time-dependent residual variability, we complete equation (\ref{Mixed0}) by the following specification of the residual part:
\begin{equation}
    \epsilon_{ij}(t_{ij}) \sim \mathcal{N}(0,\sigma_i^2(t_{ij})) \hspace{3mm} \text{with} \hspace{3mm} \log(\sigma_i(t_{ij}))  = O_{ij}^{\top} \mu+M_{ij}^{\top} \tau_{i}
 \label{Mixed1}
\end{equation}
with $O_{ij}$ and $M_{ij}$ two vectors of explanatory variables for subject $i$ at visit $j$, respectively associated with the fixed-effect vector $\mu$, and the subject-specific random-effect vector $\tau_i$, such as 
\begin{equation}
	\begin{pmatrix}
		b_{i} \\
		\tau_{i}
	\end{pmatrix}
	\sim
	\mathcal{N}\left(
	\begin{pmatrix}
		0 \\
		0
	\end{pmatrix},
	\Sigma_b = 
	\begin{pmatrix}
		\Sigma_{b} & \Sigma_{\tau b}\\
		\Sigma_{\tau b}^{\top} & \Sigma_{\tau}
	\end{pmatrix}
	\right)
	\label{CovB}
\end{equation}
As conventionally assumed in practice, the random effects $b_i$ can be considered independent of the residual component by setting $\Sigma_{\tau b} =0$.\\

\subsubsection{LSMM distinguishing within and between visits variabilities}
In some studies, several measures of the marker are collected at each measurement time. For instance, in medical studies, 2 or 3 measures of blood pressure are generally collected at each visit and intra-visit variability can be an interesting indicator \citep{courcoul2_2024}.
To define the LSMM distinguishing within and between visits variabilities, let us consider an additional level in longitudinal data, grouping repeated measurements according to measurement time. For each subject $i$, $(i = 1,\ldots,N)$, $Y_{ijl}$ is the marker value for measure $l$ $(l = 1,\ldots,n_{ij})$, at visit $j$ $(j = 1,\ldots,n_i)$ and time $t_{ij}$. For visit $j$, subject $i$ can have $n_{ij}$ measurements of the longitudinal marker. $Y_i$ is a vector of dimension $\sum_{j=1} ^{n_{i}} n_{ij}$ containing all marker measurements for the subject $i$. 

The LSMM model decomposing within and between visits individual residual variance is:
\begin{equation}
\left\{
    \begin{array}{ll}
         Y_{ijl} =  \widetilde{Y}_i(t_{ij}) + \epsilon_{ij} + \nu_{ijl} = X_{ij}^{\top} \beta+Z_{ij}^{\top} b_{i}+\epsilon_{ij} + \nu_{ijl}, \\
        \epsilon_{ij} \sim \mathcal{N}(0,\sigma_i^2) \hspace{4mm} \text{with} \hspace{3mm} \log(\sigma_i)  = \mu_\sigma + \tau_{\sigma i},\\
        \nu_{ijl} \sim \mathcal{N}(0,\kappa_i^2) \hspace{3mm} \text{with} \hspace{3mm} \log(\kappa_i)  = \mu_\kappa + \tau_{\kappa i},\\
    \end{array}
\right.  
 \label{Mixed2}
\end{equation}
with $X_{ij}$ and $Z_{ij}$ two vectors of explanatory variables for subject $i$ at visit $j$, respectively associated with the fixed-effect vector $\beta$ and the subject-specific random-effect vector $b_i$. The parameter $\mu_\sigma$ and $\mu_\kappa$ are the fixed intercepts for the between-visits and within-visit variances, respectively. The subject-specific random-effect $b_i$ and $\tau_i = (\tau_{\sigma i},\tau_{\kappa i})^\top$ are assumed to be Gaussian as presented in equation (\ref{CovB}).\\ 

In the following, we denote $r_i$ the vector of all random effects. For the standard linear mixed model, $r_i = b_i$ and for LSMMs $r_i=(b_i^\top,\tau_i^\top)^\top$.\\

\subsection{The location-scale joint models (LSJM)}

A joint model is composed of two submodels: a linear mixed model as defined by equation (\ref{Mixed0}), or a LSMM defined either by equation (\ref{Mixed1}) or by equation (\ref{Mixed2}) and a survival model. Three main survival processes represented in Figure \ref{DiffRisks} are considered and detailed in this section. Each of the defined joint model falls within the framework of shared random effect models where the time-to-event may depend on the expected value of the marker, and/or the trajectory slope, and/or its variability. 

Whatever the survival pattern, the transition intensities from state $k \in \{0,1\}$ to $l \in \{1,2\}$ are defined by:
\begin{equation}
	\lambda_{i}^{kl}(t|b_i, \tau_i)=\lambda_{0}^{kl}(t) \exp \left(W_{i}^{kl^\top} \gamma^{kl}+g_y^{kl}(b_i,t)^\top \alpha_b^{kl} + g_\tau^{kl}(\tau_i,t)^\top \alpha_\tau^{kl} \right)
	\label{Surv3}
\end{equation}
with $\lambda_{0}^{kl}(t)$ the baseline risk function, $W_{i}^{kl}$ a vector of baseline covariates associated with the regression coefficient $\gamma^{kl}$; $\alpha_b^{kl}$ is a vector of the regression coefficients associated with the function $g_y^{kl}$, possibly multi-dimentional, which measures the association between the risk and the mean evolution of $Y$. This function can be defined in the package by
\begin{itemize}
	\item the current value: $g_y^{kl}(b_i,t) = \tilde{y}_i(t)$
	\item the current slope: $g_y^{kl}(b_i,t) = \tilde{y}'_i(t)=\frac{\partial \widetilde{y_i}(t)}{\partial t}$
	\item both current value and slope: $g_y^{kl}(b_i,t) = \left(\widetilde{y_i}(t),\tilde{y}'_i(t)\right)^\top$
	\item the random effects: $g_y^{kl}(b_i,t) = b_i$
\end{itemize}
The vector $\alpha_\tau^{kl}$ includes the regression coefficients associated with the function $g_\tau^{kl}$ which measures the association between the event risk and the evolution of the residual variance of the marker. If a standard linear mixed model with an homogeneous variability is estimated for the longitudinal part then $g_\tau^{kl}(\tau_i,t) = 0$. In the case of a LSMM, two shared associations resulting from the residual variability can be considered: 
\begin{itemize}
	\item $g_\tau^{kl}(\tau_i,t) = \sigma_i(t)$ if a covariate and/or time-dependent variability is considered as in equation (\ref{Mixed1}). Thus the risk depends on the current value of the variability $\sigma_i(t)$ of the marker.
	\item $g_\tau^{kl}(\tau_i,t) = (\sigma_i , \kappa_i)^\top$ if within and between visits variabilities are considered as in equation (\ref{Mixed2}). The vector $\alpha_\tau^{kl}$ thus includes $\alpha_{\sigma}^{kl}$ and $\alpha_{\kappa}^{kl}$, the regression coefficients associated with the between-visits residual variability $\sigma_i$ and the within-visit residual variability $\kappa_i$, respectively. Note that each intensity transition can be adjusted on both variabilities or only on one of them.
\end{itemize}

For the baseline risk function, different parametric forms can be considered:
\begin{itemize}
	\item Exponential specified by $\lambda_0^{kl}(t) = \exp(\alpha_0^{kl})$
	\item Weibull specified by $\lambda_0^{kl}(t) = \zeta^{{kl}^2}t^{\zeta^{{kl}^2}-1}\exp(\alpha_0^{kl})$
	\item Gompertz specified by $\lambda_0^{kl}(t) = \kappa_1^{{kl}^2}\exp(\kappa_2^{kl}t)$
	\item Cubic B-splines, with $Q$ knots, specified by $\lambda_{0}^{kl}(t) = \exp\left(\sum_{q=1}^{Q+4} \eta_{q}^{kl} B_q(t,\nu)\right)$ where $B_q(t,\nu)$ is the q-th basis function of B-splines with the knot vector $\nu$ and $\eta_{q}^{kl}$ is the associated parameter to be estimated. 
\end{itemize}

In the following, we will describe the different survival frameworks.\\

\subsubsection{A single event: a proportional hazard model} \label{Single}

In the simplest case we consider only a single event (see Figure \ref{DiffRisks}, left panel). In that case, there is only one transition intensity between state ($0$) and state ($1$). We denote $T_i = \min(T^*_{i1},C_i)$ the observed time with $T^*_{i1}$ the real time for the event and $C_i$ the censoring time for the $i$th individual. Censoring time is assumed to be non-informative (i.e. independent from the time to event given characteristics of marker trajectories and covariates). We then denote $\delta_i \in \{0,1\}$ the individual event indicator such as $\delta_i = 1$ if $T^*_{i1} \leq C_i$ and $\delta_i = 0$ otherwise.

\subsubsection{Competing events: cause-specific model}

Considering two competing events (see Figure \ref{DiffRisks}, central panel), the observed time is then defined by $T_i = \min(T^*_{i1},T^*_{i2},C_i)$ with $T^*_{ik}$ the real time for the event $k$ $(k=1,2)$ and $C_i$ the censoring time for the $i$th individual. We denote also $\delta_i \in \{0,1,2\}$ the individual event indicator such as $\delta_i = k$ if $T_i = T^*_{ik}$ for $k \in \{1,2\}$ and $\delta_i = 0$ otherwise. In this case, the cause-specific proportional hazard model is defined using equation (\ref{Surv3}) for transition intensities from state ($0$) to ($1$) or ($2$).

\subsubsection{Semi-competing event: illness-death model}
\begin{figure}[!ht]
\centering
\resizebox{1\textwidth}{!}{%
\begin{tikzpicture}
\tikzstyle{every node}=[font=\Large]
\draw  (1.75,10.75) circle (0.5cm) node {\LARGE 0} ;
\draw  (5.5,10.75) circle (0.5cm) node {\LARGE 1} ;
\draw [->, >=Stealth] (2.25,10.75) -- (5,10.75)node[pos=0.5, fill=white]{$\lambda_i^{01}(t)$};
\draw  (8.5,9.25) circle (0.5cm) node {\LARGE 0} ;
\draw  (11.25,11) circle (0.5cm) node {\LARGE 1} ;
\draw  (11.25,7.5) circle (0.5cm) node {\LARGE 2} ;
\draw [->, >=Stealth] (9,9.5) -- (10.75,10.75)node[pos=0.5, fill=white]{$\lambda_i^{01}(t)$};
\draw [->, >=Stealth] (9,9) -- (10.75,7.75)node[pos=0.5, fill=white]{$\lambda_i^{02}(t)$};
\draw  (14.5,10.75) circle (0.5cm) node {\LARGE 0} ;
\draw  (17.75,10.75) circle (0.5cm) node {\LARGE 1} ;
\draw  (17.75,8) circle (0.5cm) node {\LARGE 2} ;
\draw [->, >=Stealth] (15,10.75) -- (17.25,10.75)node[pos=0.5, fill=white]{$\lambda_i^{01}(t)$};
\draw [->, >=Stealth] (17.75,10.25) -- (17.75,8.5)node[pos=0.5, fill=white]{$\lambda_i^{12}(t)$};
\draw [->, >=Stealth] (14.75,10.25) -- (17.25,8.25)node[pos=0.5, fill=white]{$\lambda_i^{02}(t)$};
\node [font=\Large] at (2.75,12.5) {Single event};
\node [font=\Large] at (9.75,12.5) {Competing events};
\node [font=\Large] at (16.75,12.5) {Semi-competing events};
\draw [ dashed] (0.75,13.25) rectangle  (6.25,6);
\draw [ dashed] (7.25,13.25) rectangle  (13,6);
\draw [ dashed] (13.5,13.25) rectangle  (20,6);
\end{tikzpicture}
}%
\caption{Graphical representations of the three different patterns for the survival part.}
\label{DiffRisks}
\end{figure}
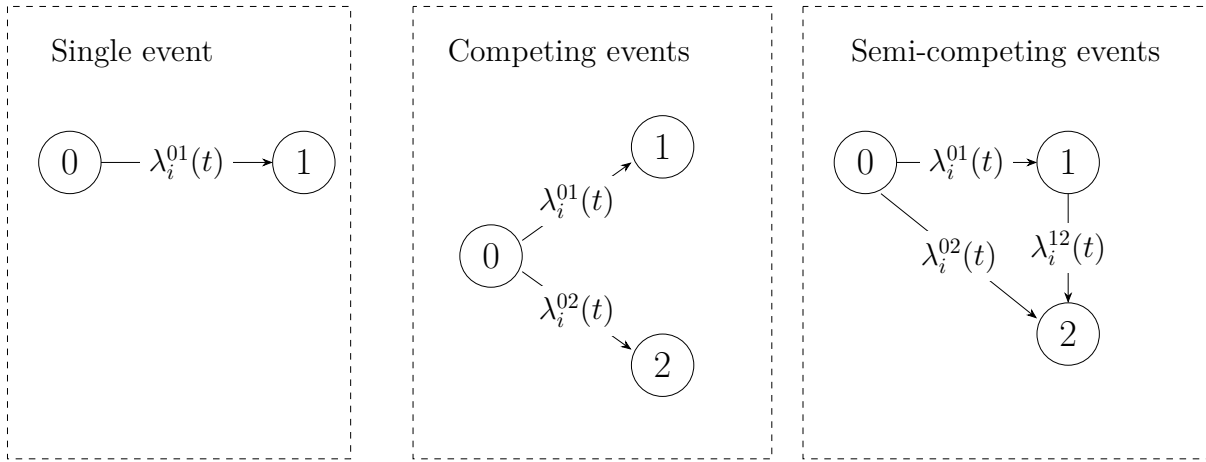

The illness-death model is a multi-state model (see Figure \ref{DiffRisks}, right panel) which describes the transitions from an initial state (e.g., alive and disease-free) to an absorbing state (e.g death) either directly or via an intermediate state (e.g. disease such as dementia). In some applications, for example in the study of dementia and death (state $(0)$ for healthy, state $(1)$ for dementia and $(2)$ for death), the transition times from the initial state $(0)$ (healthy) to the intermediate state $(1)$ (dementia) can be interval-censored for some or all subjects. This typically occurs when the intermediate state is only checked at visit times. For instance, if a subject is diagnosed with dementia at a visit time $R$, and was free of dementia at previous visit time $L$, the onset of dementia is interval-censored between $L$ and $R$ for that subject. The occurrence of death can lead to a complex observational pattern: if a subject dies without a prior diagnosis of dementia, it is not known whether the disease onset occurred between the last visit and the time of death. We assume that time of transition to state $(1)$ could be interval-censored while time of transition to state $(2)$ is known. The vector of collected data for time-to-events is given by $D_i = (T_{0i}, L_i, R_i, \delta_i^{(1)}, T_i, \delta_i^{(2)})^\top$ where $T_{0i}$ is the time at inclusion in case of delayed entry, $L_i$ is the time at the last visit where the subject was still in state $(0)$, $R_i$ is the time at the first visit where the subject is seen in state $(1)$ (in case of no observation in state $(1)$, $R_i = +\infty$), $T_i$ is the minimum between the time of transition to state $(2)$ and the time of right censoring (e.g. the end of the follow-up), $\delta_i^{(1)} =\mathds{1}_{\left\{R_i<T_i\right\} }$ is the indicator of transition to state $(1)$ and $\delta_i^{(2)}$ is the indicator of transition to state $(2)$.\\

The transition intensities from state $k \in \{0,1\}$ to state $l \in \{1,2\}$ are defined by a proportional hazards model under the Markovian hypothesis using equation (\ref{Surv3}).

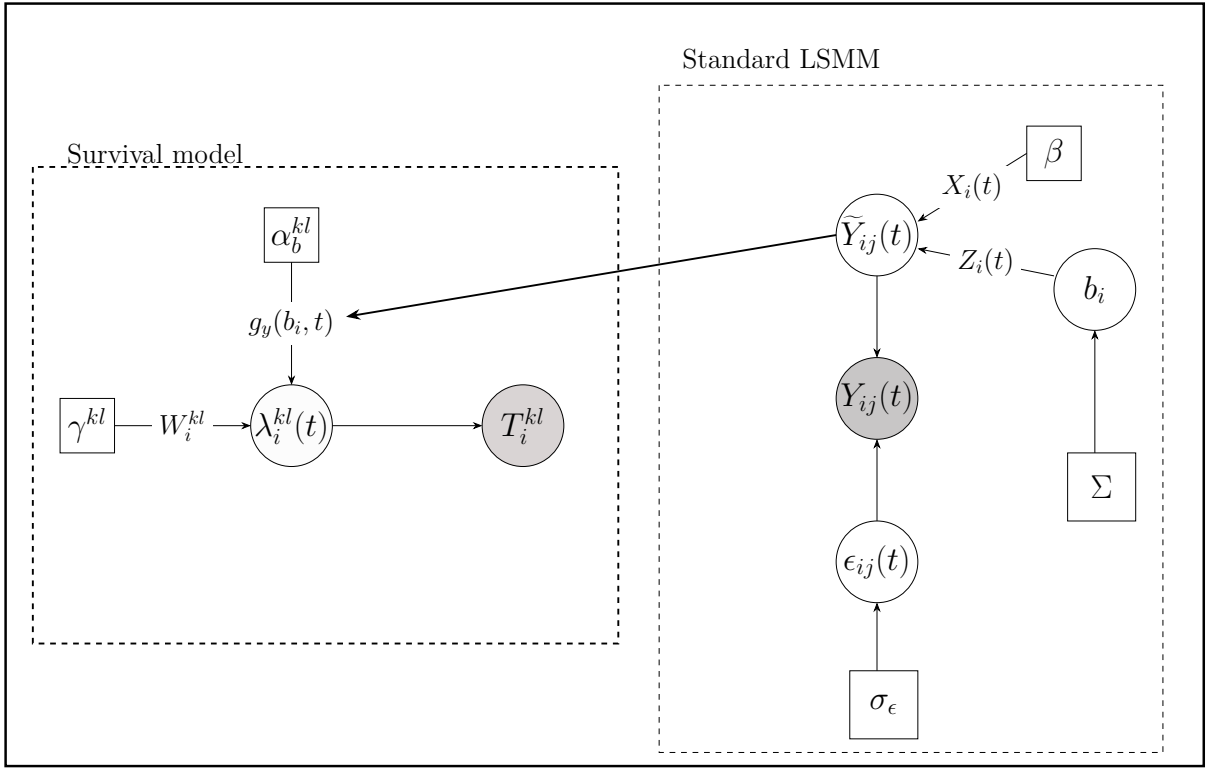
\begin{figure}[h]
\centering
\resizebox{1\textwidth}{!}{%
\begin{tikzpicture}
\tikzstyle{every node}=[font=\Large]
\draw  (13.75,11.25) rectangle  node {\LARGE $\beta$} (14.75,10.25);
\draw  (11,9.25) circle (0.75cm) node {\LARGE $\widetilde{Y}_{ij}(t)$} ;
\draw [->, >=Stealth] (13.75,10.75) -- (11.75,9.5)node[pos=0.5, fill=white]{$X_i(t)$};
\draw  (15,8.25) circle (0.75cm) node {\LARGE $b_i$} ;
\draw  (14.5,5.25) rectangle  node {\LARGE $\Sigma$} (15.75,4);
\draw [ fill={rgb,255:red,200; green,198; blue,198} ] (11,6.25) circle (0.75cm) node {\LARGE $Y_{ij}(t)$} ;
\node [font=\Large] at (9.25,12.5) {Standard LSMM};
\draw [ fill={rgb,255:red,252; green,252; blue,252} ] (0.25,5.75) circle (0.75cm) node {\LARGE $\lambda_i^{kl}(t)$} ;
\draw [ fill={rgb,255:red,217; green,212; blue,212} ] (4.5,5.75) circle (0.75cm) node {\LARGE $T_i^{kl}$} ;
\draw [->, >=Stealth] (1,5.75) -- (3.75,5.75);
\draw  (-0.25,9.75) rectangle  node {\LARGE $\alpha_b^{kl}
$} (0.75,8.75);
\draw [->, >=Stealth] (0.25,8.75) -- (0.25,6.5)node[pos=0.5, fill=white]{$g_y(b_i,t)$};
\draw  (-4,6.25) rectangle  node {\LARGE $\gamma^{kl}$} (-3,5.25);
\draw [->, >=Stealth] (-3,5.75) -- (-0.5,5.75)node[pos=0.5, fill=white]{$W_i^{kl}$};
\node [font=\Large] at (-2.25,10.75) {Survival model};
\node [font=\LARGE] at (-3,14) {Standard LSJM};
\node [font=\Large] at (11.25,7.5) {};
\node [font=\Large] at (11.25,7.5) {};
\draw [line width=1pt, ->, >=Stealth] (10.25,9.25) -- (1.25,7.75);
\draw [ line width=1pt , dashed] (-4.5,10.5) rectangle  (6.25,1.75);
\draw  (11,3.25) circle (0.75cm) node {\LARGE $\epsilon_{ij}(t)$} ;
\node [font=\Large, color={rgb,255:red,55; green,43; blue,151}] at (10.25,1.75) {};
\node [font=\Large, color={rgb,255:red,55; green,43; blue,151}] at (10.25,1.75) {};
\node [font=\Large, color={rgb,255:red,55; green,43; blue,151}] at (10.25,1.75) {};
\node [font=\Large, color={rgb,255:red,31; green,63; blue,224}] at (10.25,1.75) {};
\draw [->, >=Stealth] (11,8.5) -- (11,7);
\draw [->, >=Stealth] (14.25,8.5) -- (11.75,9)node[pos=0.5, fill=white]{$Z_i(t)$};
\draw [->, >=Stealth] (11,4) -- (11,5.5);
\draw [ dashed] (7,12) rectangle  (16.25,-0.25);
\draw [ line width=1.3pt ] (-5,13.5) rectangle (17,-0.5);
\draw  (10.5,1.25) rectangle  node {\LARGE $\sigma_\epsilon$} (11.75,0);
\draw [->, >=Stealth] (11,1.25) -- (11,2.5);
\draw [->, >=Stealth] (15,5.25) -- (15,7.5);
\end{tikzpicture}

}%
\caption{Graphical representation of the linear mixed model (right box) and the joint model (global box). The survival part could be defined either by only one cause-specific risk function (for one event), two cause-specific risk functions (for two competing events) or an illness-death model (for two semi-competing events). Circles denote random variables (white circles denote latent variables) and square boxes denote unknown parameters. $X_i(t)$ and  $Z_i(t)$ are the design matrices for individual $i$ at time $t$.}
\label{StdLS}
\end{figure}

\begin{figure}[h]
\centering
\resizebox{1\textwidth}{!}{%
\begin{tikzpicture}
\tikzstyle{every node}=[font=\Large]
\tikzstyle{every node}=[font=\Large]
\draw  (13.75,11.25) rectangle  node {\LARGE $\beta$} (14.75,10.25);
\draw  (11,9.25) circle (0.75cm) node {\LARGE $\widetilde{Y}_{ijl}(t)$} ;
\draw [->, >=Stealth] (13.75,10.75) -- (11.75,9.5)node[pos=0.5, fill=white]{$X_i(t)$};
\draw  (15,8.25) circle (0.75cm) node {\LARGE $b_i$} ;
\draw  (18,7) rectangle  node {\LARGE $\Sigma$} (19.25,5.75);
\draw [ fill={rgb,255:red,200; green,198; blue,198} ] (11,6.25) circle (0.75cm) node {\LARGE $Y_{ijl}(t)$} ;
\node [font=\Large] at (11.25,12.5) {Within and Between visits LSMM};
\draw [ fill={rgb,255:red,252; green,252; blue,252} ] (0.25,5.75) circle (0.75cm) node {\LARGE $\lambda_i^{kl}(t)$} ;
\draw [ fill={rgb,255:red,217; green,212; blue,212} ] (4.5,5.75) circle (0.75cm) node {\LARGE $T_i^{kl}$} ;
\draw [->, >=Stealth] (1,5.75) -- (3.75,5.75);
\draw  (-0.25,9.75) rectangle  node {\LARGE $\alpha_b^{kl}
$} (0.75,8.75);
\draw [->, >=Stealth] (0.25,8.75) -- (0.25,6.5)node[pos=0.5, fill=white]{$g_y(b_i,t)$};
\draw  (-4,6.25) rectangle  node {\LARGE $\gamma^{kl}$} (-3,5.25);
\draw [->, >=Stealth] (-3,5.75) -- (-0.5,5.75)node[pos=0.5, fill=white]{$W_i^{kl}$};
\node [font=\Large] at (-2.25,10.75) {Survival model};
\node [font=\LARGE] at (-1.3,14) {Within and Between visits LSJM};
\node [font=\Large] at (11.25,7.5) {};
\node [font=\Large] at (11.25,7.5) {};
\draw  (-0.25,3) rectangle  node {\LARGE $\alpha_\tau
^{kl}
$} (0.75,2);
\draw [line width=1pt, ->, >=Stealth] (0.25,3) -- (0.25,5)node[pos=0.5, fill=white]{$g_\tau(\tau_i,t)$};
\draw [line width=1pt, ->, >=Stealth] (10.25,9.25) -- (1.25,7.75);
\draw [ line width=1pt , dashed] (-5,10.25) rectangle  (5.75,1.5);
\node [font=\Large, color={rgb,255:red,154; green,154; blue,154}] at (16.5,5) {Scale part};
\node [font=\Large, color={rgb,255:red,154; green,154; blue,154}] at (17.75,11) {Location part};
\draw  (9.75,3.75) circle (0.75cm) node {\LARGE $\epsilon_{ij}(t)$} ;
\draw  (13.5,3.25) circle (0.75cm) node {\LARGE $\nu_{ijl}(t)$} ;
\draw [line width=1pt, ->, >=Stealth] (9.75,4.5) -- (10.5,5.75);
\draw  (7,-3.5) rectangle  node {\LARGE $\mu_\epsilon$} (8,-4.5);
\draw  (13,0) rectangle  node {\LARGE $\mu_\nu
$} (14,-1);
\draw  (20.5,0.5) circle (0.75cm) node {\LARGE $\tau_i$} ;
\draw  (11.75,-4) circle (0.75cm) node {\LARGE $\tau_{\epsilon i}$} ;
\draw  (16.25,-0.5) circle (0.75cm) node {\LARGE $\tau_{\nu i}$} ;
\node [font=\Large, color={rgb,255:red,55; green,43; blue,151}] at (10.25,1.75) {};
\node [font=\Large, color={rgb,255:red,55; green,43; blue,151}] at (10.25,1.75) {};
\node [font=\Large, color={rgb,255:red,55; green,43; blue,151}] at (10.25,1.75) {};
\node [font=\Large, color={rgb,255:red,31; green,63; blue,224}] at (10.25,1.75) {};
\draw [->, >=Stealth] (13.25,4) -- (11.75,5.75);
\draw [->, >=Stealth] (16.25,0.25) -- (16.25,1);
\draw [->, >=Stealth] (11,8.5) -- (11,7);
\draw [->, >=Stealth] (14.25,8.5) -- (11.75,9)node[pos=0.5, fill=white]{$Z_i(t)$};
\draw [dashed] (11.75,6.25) -- (18,6.25);
\draw  (16,1.75) circle (0.75cm) node {\LARGE $\sigma_{\nu i}$} ;
\draw [->, >=Stealth] (15.5,2.25) -- (14.25,3);
\draw [->, >=Stealth] (14,0) -- (15.5,1.25);
\draw  (9.75,0.75) circle (0.75cm) node {\LARGE $\sigma_{\epsilon i}$} ;
\draw [->, >=Stealth] (9.75,1.5) -- (9.75,3);
\draw [->, >=Stealth] (8,-3.5) -- (9.5,0);
\draw [->, >=Stealth] (11.75,-3.25) -- (10,0);
\draw [->, >=Stealth] (20,0) -- (12.5,-3.75);
\draw [->, >=Stealth] (19.75,0.5) -- (17,-0.25);
\draw [->, >=Stealth] (19.25,5.75) -- (20.5,1.25);
\draw [->, >=Stealth] (18,7) -- (15.75,8);
\draw [->, >=Stealth] (9,0.75) -- (1,3.75);
\draw [->, >=Stealth] (15.25,1.75) -- (1,4);
\draw [ dashed] (6.5,11.75) rectangle  (21.5,-5);
\draw [dashed] (6.5,6.25) -- (10.25,6.25);
\draw [dashed] (19.25,6.25) -- (21.5,6.25);
\draw [ line width=1.3pt ] (-6.5,13.5) rectangle (22,-5.5);
\end{tikzpicture}
}%
\caption{Graphical representation of the location-scale mixed model (right box) and location-scale joint model (global box) distinguishing within and between visits residual variabilities. The survival part (left box) could be defined either by only one cause-specific risk function (for one event), two cause-specific risk functions (for two competing events) or three transition intensities of an illness-death model (for two semi-competing events). Circles denote random variables, the white ones denote latent variables, and square boxes denote unknown parameters. $X_i(t)$ and  $Z_i(t)$ are the design matrices for individual $i$ at time $t$.}
\label{DoubleVarMod}
\label{fig:my_label}
\end{figure}
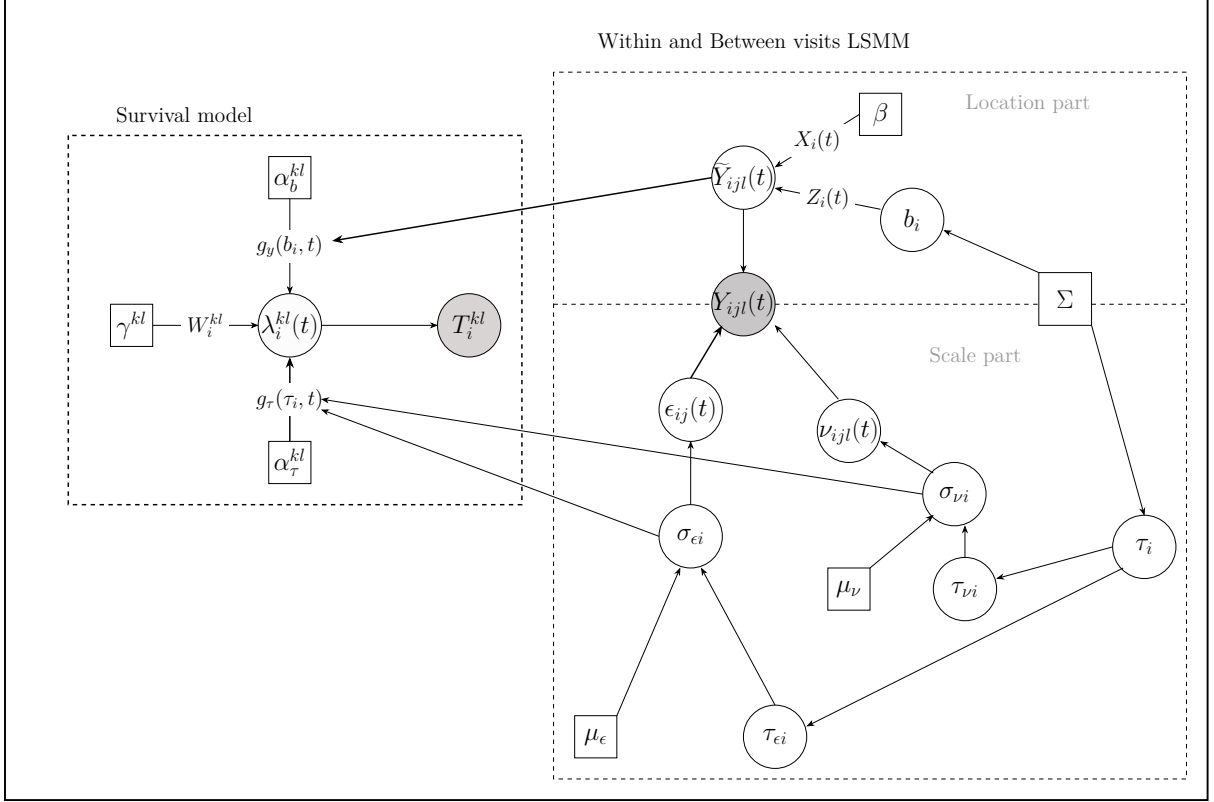

To get a better view of the dependence structure, Figures (\ref{StdLS}), (\ref{TimeDepMod}) and (\ref{DoubleVarMod}) illustrate the graphical representations of the LSJMs for the three different LSMMs (standard, time-dependent or combining between and within visits), respectively. For readers specifically interested in the LSMM, refer only to the box on the right corresponding to the representation of the LSMM. \\

\begin{figure}[h]
\centering
\resizebox{1\textwidth}{!}{%
\begin{tikzpicture}
\tikzstyle{every node}=[font=\Large]
\tikzstyle{every node}=[font=\Large]
\draw  (13.75,11.25) rectangle  node {\LARGE $\beta$} (14.75,10.25);
\draw  (11,9.25) circle (0.75cm) node {\LARGE $\widetilde{Y}_{ij}(t)$} ;
\draw [->, >=Stealth] (13.75,10.75) -- (11.75,9.5)node[pos=0.5, fill=white]{$X_i(t)$};
\draw  (15,8.25) circle (0.75cm) node {\LARGE $b_i$} ;
\draw  (18,7) rectangle  node {\LARGE $\Sigma$} (19.25,5.75);
\draw [ fill={rgb,255:red,200; green,198; blue,198} ] (11,6.25) circle (0.75cm) node {\LARGE $Y_{ij}(t)$} ;
\node [font=\Large] at (11.25,12.5) {Time-dependent LSMM};
\draw [ fill={rgb,255:red,252; green,252; blue,252} ] (0.25,5.75) circle (0.75cm) node {\LARGE $\lambda_i^{kl}(t)$} ;
\draw [ fill={rgb,255:red,217; green,212; blue,212} ] (4.5,5.75) circle (0.75cm) node {\LARGE $T_i^{kl}$} ;
\draw [->, >=Stealth] (1,5.75) -- (3.75,5.75);
\draw  (-0.25,9.75) rectangle  node {\LARGE $\alpha_b^{kl}
$} (0.75,8.75);
\draw [->, >=Stealth] (0.25,8.75) -- (0.25,6.5)node[pos=0.5, fill=white]{$g_y(b_i,t)$};
\draw  (-4,6.25) rectangle  node {\LARGE $\gamma^{kl}$} (-3,5.25);
\draw [->, >=Stealth] (-3,5.75) -- (-0.5,5.75)node[pos=0.5, fill=white]{$W_i^{kl}$};
\node [font=\Large] at (-2.25,10.75) {Survival model};
\node [font=\LARGE] at (-1.5,14) {Time-dependent LSJM};
\node [font=\Large] at (11.25,7.5) {};
\node [font=\Large] at (11.25,7.5) {};
\draw  (-0.25,3) rectangle  node {\LARGE $\alpha_\tau
^{kl}
$} (0.75,2);
\draw [line width=1pt, ->, >=Stealth] (0.25,3) -- (0.25,5)node[pos=0.5, fill=white]{$g_\tau(\tau_i,t)$};
\draw [line width=1pt, ->, >=Stealth] (10.25,9.25) -- (1.25,7.75);
\draw [ line width=1pt , dashed] (-5,10.25) rectangle  (5.75,1.5);
\node [font=\LARGE, color={rgb,255:red,154; green,154; blue,154}] at (19.25,4.75) {Scale part};
\node [font=\Large, color={rgb,255:red,154; green,154; blue,154}] at (17.75,11) {Location part};
\draw  (11,3.25) circle (0.75cm) node {\LARGE $\epsilon_{ij}(t)$} ;
\draw  (14.5,4.25) circle (0.75cm) node {\LARGE $\tau_i$} ;
\node [font=\Large, color={rgb,255:red,55; green,43; blue,151}] at (10.25,1.75) {};
\node [font=\Large, color={rgb,255:red,55; green,43; blue,151}] at (10.25,1.75) {};
\node [font=\Large, color={rgb,255:red,55; green,43; blue,151}] at (10.25,1.75) {};
\node [font=\Large, color={rgb,255:red,31; green,63; blue,224}] at (10.25,1.75) {};
\draw [->, >=Stealth] (11,8.5) -- (11,7);
\draw [->, >=Stealth] (14.25,8.5) -- (11.75,9)node[pos=0.5, fill=white]{$Z_i(t)$};
\draw [dashed] (11.75,6.25) -- (18,6.25);
\draw [->, >=Stealth] (11,4) -- (11,5.5);
\draw [->, >=Stealth] (18,7) -- (15.75,8);
\draw [ dashed] (7.5,12) rectangle  (21,-0.75);
\draw [dashed] (7.5,6.25) -- (10.25,6.25);
\draw [dashed] (19.25,6.25) -- (21,6.25);
\draw [ line width=1.3pt ] (-5.25,13.5) rectangle (21.25,-1);
\draw  (11,0.5) circle (0.75cm)  node {\LARGE $\sigma_i(t)
$};
\draw  (14.5,1.25) rectangle  node {\LARGE $\mu
$} (15.75,0);
\draw [->, >=Stealth] (18,5.75) -- (15.25,4.5);
\draw [->, >=Stealth] (14.5,0.5) -- (11.75,0.5)node[pos=0.5, fill=white]{$O_i(t)$};
\draw [->, >=Stealth] (14.2,3.5) -- (11.75,1.3)node[pos=0.5, fill=white]{$M_i(t)$};
\draw [->, >=Stealth] (11,1.25) -- (11,2.5);
X
\draw [->, >=Stealth] (10.5,1.25) -- (1,4);
\end{tikzpicture}
}%
\caption{Graphical representation of the location-scale mixed model (right box) and location-scale joint model (global box) with time-dependent residual variance. The survival part (left box) could be defined either by only one cause-specific risk function (for one event), two cause-specific risk functions (for two competing events) or three transition intensities of an illness-death model (for two semi-competing events). Circles denote random variables, the white ones denote latent variables, and square boxes denote unknown parameters. $X_i(t), Z_i(t), O_i(t)$ and $M_i(t)$ are the design matrices for individual $i$ at time $t$.}
\label{TimeDepMod}
\end{figure}
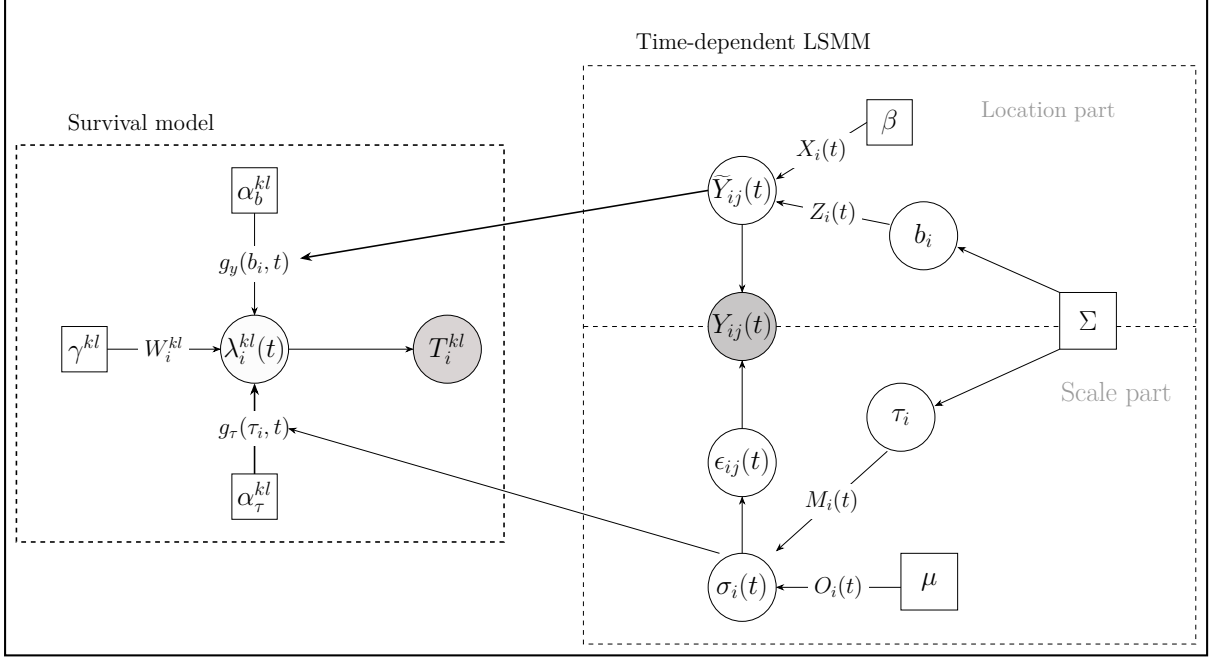

\newpage
\section{Estimation}
\label{estimationSec}
All these models are estimated within the maximum likelihood framework. We denote $\theta$ the entire vector of parameters involved in the model estimation. We consider the log-likelihood defined by $\ell(\theta) = \sum_{i=1}^N\log(\mathcal{L}_i(\theta))$ with $\mathcal{L}_i$ the individual contribution to the likelihood, which will be defined in this section for each of the models presented above.

\subsection{Individual contributions to the likelihoods}

\subsubsection{Linear location-scale mixed models}

\label{lsmm}
The individual contribution to the likelihood of a LSMM is:
\begin{equation}\label{eq:lsmm}
    \mathcal{L}_i(\theta;Y_i) =  \int f(Y_i|r_i;\theta)f(r_i;\theta)dr_i = \int \prod_{j=1}^{n_i}f(Y_{ij}|r_i;\theta)f(r_i;\theta)dr_i.
\end{equation}
For the standard linear mixed model and the LSMM with a time-dependent residual variability, $f(Y_{ij}|r_i;\theta)$ is an univariate Gaussian density and $f(r_i;\theta)$ is a multivariate Gaussian density with zero-mean and covariance $\Sigma$. Note that $r_i=b_i$ for the standard model and $r_i = (b_i,\tau_i)^\top$ for the LSMM defined by equation (\ref{CovB}).

For the LSMM with both between and within visits variabilities, $Y_{ij}$ is a vector of dimension $n_{ij}$ and $f(Y_{ij}|r_i;\theta) = f(Y_{ij}|b_i,\tau_{\sigma_i},\tau_{\kappa_i};\theta)$ is a multivariate Gaussian density with covariance matrix $\Sigma_{Y_ij|r_i} = \sigma_i^2\mathds{1}_{n_{ij}}^\top\mathds{1}_{n_{ij}} + \kappa_i^2 I_{n_{ij}}$, where $\mathds{1}_{n_{ij}}$ is the $n_{ij}$-length vector of 1 and $I_{n_{ij}}$ is the identity matrix of size $n_{ij}\times n_{ij}$.

\subsubsection{Joint models}

Let $D_i$ be the survival data vector. Joint modeling is based on the assumption of independence between $D_i$ and $Y_i$ conditional on random effects. Assuming also independent censoring, the individual contribution to the likelihood for the LSJM is defined by:
\begin{equation}
\mathcal{L}_i(\theta;Y_i,D_i) =  \int p(Y_i, D_i|r_i;\theta) f(r_i;\theta)dr_i = \int f(Y_i|r_i;\theta)f(r_i;\theta)f(D_i|r_i;\theta)dr_i
\end{equation}
where $f(Y_i|r_i;\theta)$ and $f(r_i;\theta)$ are already defined for equation \ref{eq:lsmm} according to the LSMM sub-model considered and $f(D_i|r_i;\theta)$ varies depending on the survival model considered.\\


\noindent \textbf{For a single or two competing events.}
By distinguishing a single event from two competing events by $K=1$ and $K=2$, respectively, we have:
\begin{equation}
    f(D_i|r_i;\theta) = \exp\left(-\sum_{k=1}^K \Lambda_{i}^{0k}(T_i|r_i;\theta)\right)\prod_{k=1}^{K}\lambda_{i}^{0k}(T_i|r_i;\theta)^{\mathds{1}_{\delta_i=k}}
\end{equation}
with $\Lambda_{i}^{0k}(T_i|r_i;\theta)$ the cumulative risk function given by:
\begin{equation}\label{eq:cum_risk}
    \Lambda_{i}^{0k}(T_i|r_i;\theta) = \int_0^{T_i}\lambda_{i}^{0k}(u|r_i;\theta)du
\end{equation}

\begin{figure}[t!]
    \centering
    \includegraphics[trim = 7.5cm 2cm 1cm 3cm, clip = true, scale = 0.7]{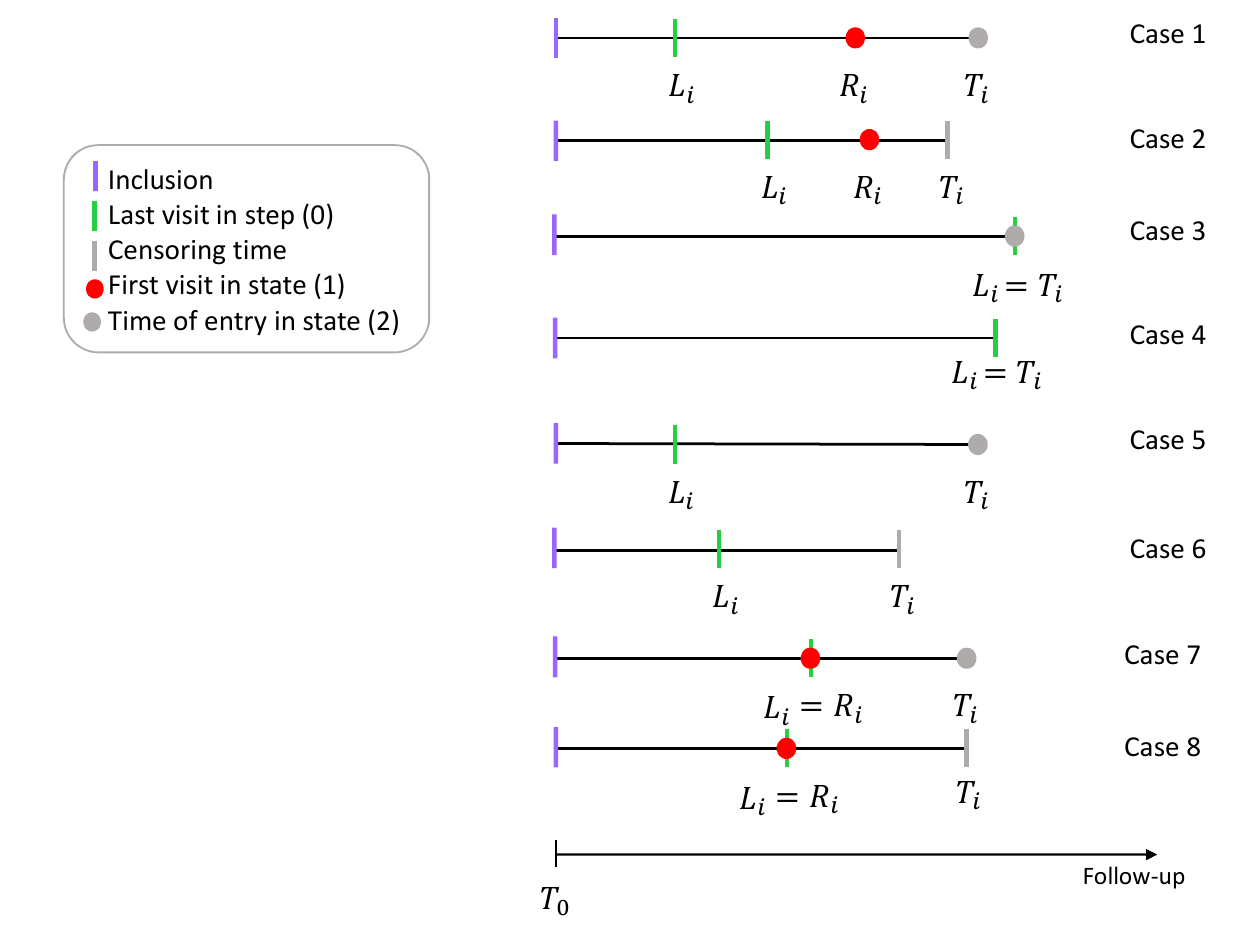}
    \caption{Possible patterns of observation for interval censored semi-competing events.}
    \label{FigCI}
\end{figure}

\noindent \textbf{For an illness-death model.}
In case of interval-censored semi-competing events and illness-death model, $f(D_i|r_i;\theta)$ depends on the subject trajectory as illustrated on Figure \ref{FigCI}.
To present the different definitions of $f(D_i|r_i;\theta)$ according to subjects observations, we deliberately omit the conditioning on random effects and parameters for ease of notations:
\begin{description}
\item[Cases 1 and 2:] subject in state $(0)$ until $L_i$, observed in state $(1)$ at $R_i$, remained in state $(1)$ until $T_i$, and either entered in state $(2)$ (case 1) or censored at $T_i$ (case 2): 
\begin{eqnarray*}
f(T_{0i}, L_i, R_i, \delta_i^{(1)} = 1, T_i, \delta_i^{(2)}) &=& \int_{L_i}^{R_i} e^{-\Lambda_{i}^{01}(u) - \Lambda_{i}^{02}(u)}\lambda_{i}^{01}(u)e^{-(\Lambda_{i}^{12}(T_i)-\Lambda_{i}^{12}(u))}\lambda_{i}^{12}(T_i)^{\delta_i^{(2)}}du
\end{eqnarray*}
\item[Cases 3 and 4:] subject in state $(0)$ until $T_i$ ($T_i = L_i$), and either entered in state $(2)$ (case 3) or censored at $T_i$ (case 4): 
\begin{eqnarray*}
f(T_{0i}, L_i, R_i, \delta_i^{(1)} = 0, T_i, \delta_i^{(2)}) &=& e^{-\Lambda_{i}^{01}(T_i) - \Lambda_{i}^{02}(T_i)}\lambda_{i}^{02}(T_i)^{\delta_i^{(2)}}
\end{eqnarray*}
\item[Cases 5 and 6:] subject in state $(0)$ at his/her last visit $L_i < T_i$, and either entered in state $(2)$ (case 5) or censored at $T_i$ (case 6):
\begin{eqnarray*}
f(T_{0i}, L_i, R_i, \delta_i^{(1)} = 0, T_i, \delta_i^{(2)}) &=& e^{-\Lambda_{i}^{01}(T_i) - \Lambda_{i}^{02}(T_i)}\lambda_{i}^{02}(T_i)^{\delta_i^{(2)}} +\\
&&\int_{L_i}^{T_i} e^{-\Lambda_{i}^{01}(u) - \Lambda_{i}^{02}(u)}\lambda_{i}^{01}(u)e^{-(\Lambda_{i}^{12}(T_i)-\Lambda_{i}^{12}(u))}\lambda_{i}^{12}(T_i)^{\delta_i^{(2)}}du
\end{eqnarray*}
The likelihood accounts for the two possible trajectories of this subject: either the subjects remains in state ($0$) until death or censoring at $T_i$ or he/she has an unobserved transition to state $(1)$ between $L_i$ and $T_i$.
\item[Cases 7 and 8:] subject observed in state $(1)$ at $L_i = R_i$ (exact date of transition observed), remained in state $(1)$ until $T_i$, and either entered in state $(2)$ (case 7) or censored at $T_i$ (case 8):
\begin{eqnarray*}
f(T_{0i}, L_i, R_i, \delta_i^{(1)} = 1, T_i, \delta_i^{(2)}) &=& e^{-\Lambda_{i}^{01}(L_i) - \Lambda_{i}^{02}(L_i)}\lambda_{i}^{01}(L_i)e^{-(\Lambda_{i}^{12}(T_i)-\Lambda_{i}^{12}(L_i))}\lambda_{i}^{12}(T_i)^{\delta_i^{(2)}}
\end{eqnarray*}
\end{description}

with $\Lambda_{i}^{12}(T_i|r_i;\theta)$ the cumulative risk function from state $1$ to state $2$ given by:
\begin{equation}\label{eq:cum_risk}
	\Lambda_{i}^{12}(T_i|r_i;\theta) = \int_0^{T_i}\lambda_{i}^{12}(u|r_i;\theta)du
\end{equation}

\noindent \textbf{Left truncation.}
In case of delayed entry, the individual contribution to the likelihood is divided by the probability to be free of any event at entry time $T_{0i}$:
\begin{equation}
    \mathcal{L}_i^{DE}(\theta;Y_i,D_i) = \frac{\mathcal{L}_i(\theta;Y_i,D_i)}{\int \exp(-\sum_{k=1}^K\Lambda_i^{0k}(T_{0i}|r_i;\theta)f(r_i;\theta)dr_i}
\end{equation}

\subsection{Computational aspects}

\subsubsection{Integral computation}

In each model, the integral over the random effects is computed by a Quasi Monte Carlo (QMC) approximation \citep{pan_quasi-monte_2007}, using deterministic quasi-random sequences. The approximation of the integral is defined by:
\begin{equation}
   \mathcal{L}_i(\theta;Y_i,D_i) \simeq \frac{1}{S} \sum_{s = 1}^S {p(Y_i,D_i|r_i^s;\theta)}
\end{equation} 
where $(r_i^1,...,r_i^S)$ are draws of a S-sample in the sobol sequel for the distribution $f(r_i;\theta)$. The sobol sequel is generated using the \pkg{spacefillr} \proglang{R} package \citep{tyler_2024}. Moreover, the cumulative risk functions defined in equation (\ref{eq:cum_risk}) are approximated using the Gauss-Kronrod quadrature with 7 or 15 points \citep{gonnet_2012}.

\subsubsection{Optimization algorithm}

Parameter estimation is performed by maximizing the log-likelihood function $\ell(\theta) = \log\left( \prod_{i=1}^N 
\mathcal{L}_i(\theta;Y_i,D_i)\right)$. This optimization is carried out using the \pkg{marqLevAlg} R-package, which applies the Marquardt-Levenberg algorithm \citep{philipps_2021}, a robust variant of the Newton-Raphson algorithm \citep{levenberg_1944, marquardt_1963}. This algorithm iteratively refines the parameter estimates $\theta$ until convergence is reached. Strict convergence criteria are used, based on the stability of both the parameters and the log-likelihood function, along with the relative distance to the maximum, calculated from the first and second derivatives of the log-likelihood. The convergence is achieved when $\frac{\nabla\left(\ell\left(\theta^{(l)}\right)\right)\left(H\left(\theta^{(l)}\right)\right)^{-1} \nabla\left(\ell\left(\theta^{(l)}\right)\right)}{m}<\varepsilon_d$ at an iteration $l$, with $m$ the number of parameters, $\varepsilon_d$ a predefined threshold, $H$ the Hessian matrix and $\nabla$ the gradient function.\\

The estimation is performed using the Cholesky decomposition of the covariance matrix of random effects to ensure positive-definite matrix. The inverse of the Hessian matrix is computed by finite difference to obtain the variances of the estimates. The variances of the estimated parameters of the covariance matrix in its natural scale are obtained using the Delta-Method. Let denote $C$ the Cholesky matrix with elements $l_{ij}$ and $\Sigma$ the covariance matrix with elements $\sigma_{ij}$ such as $\Sigma = CC^\top$. Given $Cov(\hat{l}_{ij},\hat{l}_{km})$, $Cov(\hat{\sigma}_{ij},\hat{\sigma}_{km})$ is equal to 
\begin{eqnarray*}
    Cov(\hat{\sigma}_{ij},\hat{\sigma}_{km})
    &=&\sum_{t=1}^{\min(i,j)}\sum_{s=1}^{\min(k,m)} \left[ \hat{l}_{jt}\hat{l}_{ms} Cov(\hat{l}_{it},\hat{l}_{ks}) + 
    \hat{l}_{jt}\hat{l}_{ks} Cov(\hat{l}_{it},\hat{l}_{ms})   \right. \\
    & & \left. \qquad\qquad
    + \hat{l}_{it}\hat{l}_{ms} Cov(\hat{l}_{jt},\hat{l}_{ks}) + 
    \hat{l}_{it}\hat{l}_{ks} Cov(\hat{l}_{jt},\hat{l}_{ms}) \right]
\end{eqnarray*} 

\subsubsection{Initialisation}\label{init}

In each estimation function, it is necessary to provide a set of parameter initial value. 
For estimating LSMM, the initial values of the fixed effects are obtained by estimating a standard linear regression model. The cholesky of the covariance matrix $\Sigma$ of random effects is initialised by the identity matrix. Then the estimation of the LSMM is performed using $S1$ (e.g $S1=500$ or $S1=1000$) QMC draws until convergence is achieved.
For estimating LSJM, the initial values of the location-scale mixed sub-model are the estimates obtained by fitting the LSMM alone. For the survival part, the values of the baseline risk function and the regression coefficients for the covariates $W$ are initialised using an estimation of a parametric survival model. Then, the association parameters, $\alpha_b$ and $\alpha_\tau$ are initialised at 0. For the case of an illness-death model with interval censoring, the joint model is first estimated without handling interval censoring, taking the middle of the interval as time of transition to state $(1)$ and assuming that subjects who go to state $(2)$ before being observed in state $(1)$ directly made the transition $(0)$ to $(2)$.




\subsubsection{Strategy}
For both LSMM and LSJM models, the estimation can be performed with a relatively small number $S1$ of QMC draws (e.g. $S1 = 500$ or $S1 = 1000$) to obtain estimated parameters without bias. However, as we have previously shown \citep{courcoul_SIM_2025,courcoul2_2024}, the estimated variances of parameters could be biased. We have therefore proposed to perform a second step for precision improvement, then few additional iterations are performed with a higher number $S2>S1$ of QMC draws (e.g. $S2 = 5000$ or $S2 = 10000$) until the Hessian matrix is invertible.\\

The choice of the number of QMC draws $S1$ and $S2$ has a notable effect on computational time. Therefore, for model selection, we suggest that users compare different models using the results of step 1 (using likelihood or information criteria) with a small value for $S1$, and $S2 = NULL$ and perform step 2, which entails a larger number of QMC draws, solely for the final selected model. Finally, we recommend to increase the number of QMC draws with the number of random effects and in case of convergence issues.

\section{Post-fit computations}
\label{PostSec}
The package includes a range of post-fit analyses and computations, most of which are shared between the different estimation functions. The upcoming subsections detail these post-fit computations. 

\subsection{Predictions}

For both linear mixed models and joint models, empirical Bayes estimates of the subject-specific random effects $r_i$ can be obtained. 
They correspond to the mode of their estimated conditional posterior distribution given the data and estimated parameters, denoted by $\widetilde{r}_i = \text{argmax}_{r_i}f(r_i|Y_i, \widehat{\theta})$ and $\widetilde{r}_i = \text{argmax}_{r_i}f(r_i|Y_i,T_i,\delta_i, \widehat{\theta})$ for LSMM and LSJM, respectively. Whatever the model, they are obtained using the Marquardt-Levenberg algorithm. \\

The package allows also to compute the individual predicted trajectory corresponding to the conditional expectation given the random effects, defined by:
\begin{equation}
  \widehat{\mathbb{E}}(Y_i(t)|\widetilde{r}_i) = X_i(t)\widehat{\beta} + Z_i(t)\widetilde{b_i}.
  \label{Bayes_esti}
\end{equation}

For LSJM, the predicted cumulative hazard function can be computed for a grid of times, by plugging the empirical Bayes estimates of the random effects in the formula for the risk functions:
\begin{equation}
    \widehat{\Lambda}_i^{kl}(t|\widetilde{r}_i;\widehat{\theta}) =\int_{0}^t \widehat{\lambda}_{0}^{kl}(u;\widehat{\theta})\exp({W_i^{kl}}^\top\widehat{\gamma}^{kl} + g_y^{kl}(\widetilde{b}_i,u)^\top \widehat{\alpha}_b^{kl} +  g_\tau^{kl}(\widetilde{\tau}_i,u)^\top\widehat{\alpha}_\tau^{kl})du
\label{Surv_comp}
\end{equation}
computed using the Gauss-Kronrod approximations.

\subsection{Goodness-of-fit}

\subsubsection{Longitudinal fit}

To assess the fit of the longitudinal sub-model, we compute the predicted marker values for each individual at their respective visit times using the empirical Bayes estimates (equation (\ref{Bayes_esti})). Then it is possible to display the means of the predicted marker values and the mean of the observed measurements grouping the observations by chosen time windows.

\subsubsection{Survival fit}

To evaluate the fit of the survival submodels, it is possible to provide a plot of the mean of the predicted cumulative hazard function (equation (\ref{Surv_comp})) for each transition versus the Nelson-Aalen estimator in case of competing risks or versus a non-parametric estimator obtained using the \pkg{SmoothHazard} package \citep{touraine_2017} in case of IDM with interval-censored data.

\subsection{Dynamic predictions with LSJM }\label{DynpredSection}

Individual dynamic predictions can be computed and plotted. They are defined as the predicted probability of having event $k$ between time $s$ and $s+t$ given that the subject did not experience any event before time $s$, and knowing all marker measures collected until time $s$, denoted by $\mathcal{Y}_i(s)$, and the set of estimated parameters $\widehat{\theta}$. The prediction is defined for subject $i$ by: 
\begin{eqnarray}
   \pi_i^{0k}(s,t;\widehat{\theta})&= &P(s<T_i<s+t, \delta_i = k | T_i >s, \mathcal{Y}_i(s), \widehat{\theta}) \nonumber \\
   &= & \frac{\int \left[\int_s^{s+t} \exp{(-\sum_{c=1}^K \Lambda_{i}^{0c}(u|r_i,\widehat{\theta}))}\lambda_{i}^{0k}(u|r_i,\widehat{\theta})du \right]f(\mathcal{Y}_i(s)|r_i,\widehat{\theta})f(r_i|\widehat{\theta})dr_i}{\int\exp{(-\sum_{c=1}^K \Lambda_{i}^{0c}(s|r_i,\widehat{\theta}))}f(\mathcal{Y}_i(s)|r_i,\widehat{\theta})f(r_i|\widehat{\theta})dr_i}\nonumber\\
   & & 
   \label{pred_st}
\end{eqnarray}
where $K=2$ for illness-death model and competing events and $K=1$ for a model with a single event. Note that only the prediction from state $(0)$ to state $(1)$ can be computed for IDM whereas both transition $(0)$ to $(1)$ and $(0)$ to $(2)$ can be predicted for competing risks. \\
As in the estimation procedure, the integral over the random effects is computed by QMC approximation and the integral over time with the Gauss-Kronrod quadrature.
The 95\% confidence interval of predictions is obtained by the following Monte Carlo algorithm. For $L$ large enough and $l  = 1,...,L$ ($L$ = 1000 for instance): 
\begin{enumerate}
    \item Generate $\widetilde{\theta}^{(l)} \sim \mathcal{N}(\widehat{\theta},V(\widehat{\theta}))$ where $V(\widehat{\theta})$ is given by the inverse of the Hessian matrix at $\widehat{\theta}$;
    \item Compute $\widetilde{\pi}^{0k^{(l)}}_i(s,t;\widetilde{\theta}^{(l)})$ from equation (\ref{pred_st});
    \item Compute the 95\% confidence interval from the 2.5th and 97.5th percentiles of the L-sample of $\widetilde{\pi}^{0k^{(l)}}_i(s,t;\widetilde{\theta}^{(l)})$.\\
\end{enumerate}

The confidence interval computation can be quite time-consuming according to $L$ and the number of QMC draws used in step 2 to approximate equation (\ref{pred_st}).\\

Finally, it is also possible to derive other quantities such as the average probability of being free of any events at a given time $t$ and covariates using the following formula:
\begin{equation}
	S(t|\widehat{\theta}) =\int \exp(-\sum_{k=1}^K \Lambda_{0k}(t|r_i,\widehat{\theta})dr_i)
\end{equation}

\section{Implementation and Examples}

The package allows to estimate the models presented in section \ref{estimationSec} and provide the output of post-fit computations detailed in section \ref{PostSec}. The two estimation functions, \fct{lsmm} and \fct{lsjm}, rely on programs written in \proglang{R} and \proglang{C++}. This section describes the data used for the examples and the calls of these functions with some illustrations based on fictive data.

\color{black}
\subsection{Fictive data}
To illustrate and showcase the full scope of the functions available in the package, we will use fictive data in the examples including the following variables:
\begin{itemize}
    \item[$\bullet$] \code{Y}: a continuous variable corresponding to the longitudinal marker measurements;
    \item[$\bullet$] \code{time}: a continuous variable corresponding to the time of the measurements;
    \item[$\bullet$] \code{ID}: the identification variable of group level (for instance the individual);
    \item[$\bullet$] \code{ID.visit}: the identification variable of the visits;
    \item[$\bullet$] \code{X1}: a time-dependent qualitative or quantitative covariate;
    \item[$\bullet$] \code{W1, W2, W3}: three time-fixed qualitative or quantitative covariates.
    \item[$\bullet$] \code{event1}: a variable corresponding to the indicator of event for transition to state (1);
    \item[$\bullet$] \code{event2}: a variable corresponding to the indicator of event for transition to state (2);
    \item[$\bullet$] \code{Time\_event}: a variable corresponding to the time of the first event (or censoring) in case of joint model with a single event or with competing events;
    \item[$\bullet$] \code{Time\_entry}: a variable corresponding to the time of entry in the study (delayed entry);
    \item[$\bullet$] \code{Time\_final}: a variable corresponding to the minimum time between time-to-transition to state (2) and censoring time;
    \item[$\bullet$] \code{Time\_left}: a variable corresponding to the last observed time in state (0);
    \item[$\bullet$] \code{Time\_right}: a variable corresponding to the first observed time known in state (1);
\end{itemize}

\subsection{Estimation function lsmm}

The syntax of \fct{lsmm} is
\begin{Code}
lsmm(formFixed, formRandom, formGroup, timeVar,
     formVar = "standard", formFixedVar = NULL, formRandomVar = NULL,
     random_inter = F, random_intra = F, formGroupVisit = NULL, correlated_re = F,
     data.long, S1 = 500, S2 = 5000, nproc = 1, clustertype = "PSOCK", 
     maxiter = 100, print.info = F,file = "", epsa = 1e-04, epsb = 1e-04,
     epsd = 1e-04, binit = NULL)
\end{Code}
There are two types of arguments: those associated with the model definition and those associated with computational aspects. For the model specification, we have:
\begin{itemize}
    \item[$\bullet$] \code{formFixed}: a two-sided formula for the regression model at the population level, with the response variable $Y$ on the left-hand side of a \code{~} operator, and the covariates associated with fixed effects, separated by + operator, on the right-hand side.
    \item[$\bullet$] \code{formRandom}: a one-sided formula with the covariates associated with random effects, separated by + operator, on the right-hand side.
    \item[$\bullet$] \code{formGroup}: a one-sided formula with the variable identifying the subjects.
    \item[$\bullet$] \code{timeVar}: a character providing the name of time variable.
    \item[$\bullet$] \code{formVar}: a character defining which definition of variability is used (details below).
    \item[$\bullet$] \code{data.long}: a database containing the data in long format, meaning there are $n_i$ rows per subject, or $\sum_{j}^{n_i} n_{ij}$ when there is an additional nested level in the individual data (e.g., multiple measurements for a single time point or a follow-up window).
    \item[$\bullet$] \code{formFixedVar, formRandomVar, random\_inter, random\_intra, formGroupVisit} and \code{correlated\_re}: specified depending on the variability used. The different patterns are described below.
\end{itemize}
Regarding the arguments specific to the computational aspect, we have:
\begin{itemize}
    \item[$\bullet$] \code{S1}: an integer proving the number of QMC draws for the fisrt step.
    \item[$\bullet$] \code{S2}: an integer proving the number of QMC draws for the second step. It should be provided only if the user want to do the precision improvement step. In this case, \code{S2}$>$\code{S1}.
    \item[$\bullet$] \code{nproc}: an integer providing the number of processors to used for parallel computing.
    \item[$\bullet$] \code{clustertype}: a character indicating a clustertype supported from \fct{makeCluster}.
    \item[$\bullet$] \code{maxiter}: an integer providing the maximum number of iterations in the optimization algorithm.
    \item[$\bullet$] \code{print.info}: a boolean indicating if the outputs of each iteration should be written.
    \item[$\bullet$] \code{file}: a character giving the name of the file where the outputs of each iteration should be written.
    \item[$\bullet$] \code{epsa}, \code{epsb}, \code{epsd}: numbers providing the threshlods for convergence criteria on the parameters, the log-likelihood and the derivatives, respectively. These arguments are specific to the \fct{MarqLevAlg} function from \pkg{MarqLevAlg} package \citep{philipps_2021}.
    \item[$\bullet$] \code{binit}: a vector providing the initial value.
\end{itemize}

\subsubsection{Standard linear mixed model}
To specify a linear mixed model with an homogeneous variability, the argument \code{formVar} should be equal to \code{"standard"}. Arguments \code{formFixedVar, formRandomVar, random\_inter, random\_intra, formGroupVisit} and \code{correlated\_re} do not need to be specified.

Below are two examples of call:
\begin{Code}
lsmmStandard <- lsmm(formFixed = Y ~ time + X1, formRandom = ~ time,
                     formGroup = ~ ID, timeVar = "time", formVar = "standard", 
                     data.long = data_lsmm)
                     
lsmmStandardTime2 <- lsmm(formFixed = Y ~ time + I(time^2) + X1, 
			  formRandom = ~ time + I(time^2), formGroup = ~ ID,
			  timeVar = "time", formVar = "standard", data.long = data_lsmm)
\end{Code}
For dataset \code{data\_lsmm}, the call \code{lsmmStandard} fits a standard linear mixed model in which the dependent variable $Y$ is explained according to \code{time} and \code{X1}. Two correlated random effects are assumed for the \code{intercept} (mandatory) and \code{time}. These random effects are grouped by \code{ID}. The call \code{lsmmStandardTime2} fits the same model while additionally including \code{time}$^2$ (i.e. quadratic trend) as both a fixed and a random effect.

\subsubsection{Location-scale mixed model with time-dependent residual variability}
To specify a location-scale mixed model with subject-specific random residual variance possibly depending on time and covariates, the argument \code{formVar} should be equal to \code{"cov-dependent"}. Argument \code{formFixedVar} defines the one-sided formula with the covariates $O$ (see equation (\ref{Mixed1})) with fixed effects on the variability. Argument \code{formRandomVar} provides the one-sided formula with the covariates $M$ (see equation (\ref{Mixed1})) with random effects on the variability. Argument \code{correlated\_re} indicates if the random effects of the mean $b$ and the random effects of the variability $\tau$ should be correlated or not.

Below are two examples of call:
\begin{Code}
lsmmCovDep1 <- lsmm(formFixed = Y ~ time + X1, formRandom = ~ time,
                   formGroup = ~ ID,  timeVar = "time", 
                   formVar = "cov-dependent", formFixedVar = ~ time + X2, 
                   formRandomVar = ~ time, correlated_re = F, 
                   data.long = data_lsmm)
\end{Code}

\begin{Code}
lsmmCovDep2 <- lsmm(formFixed = Y ~ time + X1, formRandom = ~ time,
		    formGroup = ~ ID,  timeVar = "time", 
		    formVar = "cov-dependent", formFixedVar = ~ time + X2, 
		    formRandomVar = ~ 1, correlated_re = F, 
		    data.long = data_lsmm)
\end{Code}

For dataset \code{data\_lsmm}, the call \code{lsmmCovDep1} fits a location-scale mixed model in which the dependent variable $Y$ is explained according to \code{time} and \code{X1}. On the mean, two correlated random effects are assumed for the \code{intercept} and \code{time}. The residual variability is modeled according to \code{time} and \code{X2} for the fixed part and two correlated random effects, \code{intercept} and \code{time}. Random effects of the mean are supposed to be independent from the random effects of the variability. The call \code{lsmmCovDep2} fits the same location-scale mixed model but with only a random \code{intercept} for the variability (instead of both random \code{intercept} and random \code{slope}).

\subsubsection{Location-scale mixed model with between and within visits variabilities}
To specify a location-scale mixed model with between and within visits variabilities, the argument \code{formVar} should be equal to \code{"inter-intra"}. Argument \code{random\_inter} and \code{random\_intra} are boolean indicating if the between-visits variability and the within-visit variability respectively should be subject-specific or not. Argument \code{formGroupVisit} provides a one-sided formula with the visit indicator variable. Argument \code{correlated\_re} indicates if the random effects of the mean $b$ and the random effects of the variability $\tau$ should be correlated or not. Regardless of the value of \code{correlated\_re}, if both \code{random\_inter = T} and \code{random\_intra = T}, $\tau_\sigma$ and $\tau_\kappa$ are assumed to be correlated.

Below is an example of call:
\begin{Code}
lsmmInterIntra <- lsmm(formFixed = Y ~ time + X1, formRandom = ~ time,
                          formGroup = ~ ID, timeVar = "time",
                          formVar = "inter-intra", random_inter = T, 
                          random_intra = T, formGroupVisit = ~ ID.visit, 
                          correlated_re = T, data.long = data_lsmm2)
\end{Code}
For dataset \code{data\_lsmm2}, the call \code{lsmmInterIntra} fits a location-scale mixed model in which the dependent variable $Y$ is explained according to \code{time} and \code{X1}. On the mean, two correlated random effects are assumed for the \code{intercept} and \code{time}. The between-visits and within-visit variabilities are both subject-specific by the inclusion of a random effect on their intercept, and all random effects are correlated. Visits are indicated by the \code{ID.visit} variable.

\subsection{Function lsjm}
The syntax of \fct{lsjm} is
\begin{Code}
lsjm(Objectlsmm,  survival_type = c('Single', 'CR', 'IDM'),
     formSurv_01, formSurv_02 = NULL, formSurv_12 = NULL,
     sharedtype_01, sharedtype_02 = NULL, sharedtype_12 = NULL,
     hazardBase_01, hazardBase_02 = NULL, hazardBase_12 = NULL,
     delta1, delta2 = NULL, Time_T, Time_L = NULL, Time_R = NULL, Time_T0 = NULL,
     formSlopeFixed = NULL, formSlopeRandom = NULL, 
     index_beta_slope = NULL, index_b_slope = NULL, nb.knots.splines = c(1,1,1),
     nb_pointsGK = 15, S1 = 1000, S2 = 5000, binit = NULL,
     nproc = 1, clustertype = "PSOCK", maxiter = 100,
     print.info = FALSE, file = NULL, epsa = 1e-04, eps 1e-04, epsd = 1e-04)
\end{Code}
Using the following arguments:
\begin{itemize}
    \item[$\bullet$] \code{Objectlsmm} an object of the \fct{lsmm} function.
    \item[$\bullet$] \code{survival\_type} a character defining which type of survival scheme to use (see details below).
    \item[$\bullet$] \code{formSurv\_01} is a one-sided formula providing on the right-side the regression covariates associated with the proportional hazard model, separated by a + operator.
    \item[$\bullet$] \code{hazardBase\_01} a character providing the baseline hazard function, which is in \code{c("Exponential", "Weibull", "Gompertz", "Splines")}. If \code{hazardBase\_01 = "Spline"}, argument \\\code{nb.knots.splines} indicates the number of internal knots. The knots are placed at the quantiles of event times. If \code{nb.knots.splines = 1}, the knot is at the median of the event times.
    \item[$\bullet$] \code{sharedtype\_01} a vector of character(s) indicating the form(s) of the dependence structure. If the longitudinal model estimated is a standard mixed model, \code{sharedtype\_01} should be included in \code{c("value", "slope", "random effects")}. If it is a location-scale mixed model one can add \code{"variability"} and if it is a model distinguishing within from between visits variabilities one can add \code{c("variability inter", "variability intra")} if there are subject-specifics. If \code{"slope"} is included in \code{sharedtype\_01}, arguments \code{formSlopeFixed, formSlopeRandom, index\_beta\_slope, index\_b\_slope} should be provided.
    \item[$\bullet$] \code{formSlopeFixed} a one-sided formula corresponding to the derivative with respect to time of the function of fixed effects given by \code{formFixed}.
    \item[$\bullet$] \code{index\_beta\_slope} a vector of integer(s) providing the index of $\beta$ parameters from the mixed model used in the slope.
    \item[$\bullet$] \code{formSlopeRandom} a one-sided formula corresponding to the derivative with respect to time of the function of random effects given by \code{formRandom}.
    \item[$\bullet$] \code{index\_b\_slope} a vector of integer(s) indicating the index of $b$ parameters from the mixed model used in the slope.
    \item[$\bullet$] \code{delta1} a one-sided formula with on the right-side the variable corresponding to the indicator $\delta_{1i}$ of event (1 for subject experimenting the event and 0 for others).
    \item[$\bullet$] \code{Time\_T} a one-sided formula with on the right-side the variable corresponding to the time $T_i$ of event.
    \item[$\bullet$] \code{Time\_T0} a one-sided formula with on the right-side the variable corresponding to the time of entry $T_{0i}$ which must be provided in case of delayed entry.
\end{itemize}
Arguments \code{formSurv\_02, formSurv\_12, sharedtype\_02, sharedtype\_12, hazardBase\_02, hazardBase\_12, delta2, Time\_L, Time\_R} are specified similarly to previous arguments, depending on the chosen survival model as described below. Argument \code{nb\_pointsGK} provides the number of points to use for the Gauss-Kronrod approximation (either 7 or 15). The other arguments, related to the computationel aspect are the same than for \code{lsmm} function. Argument \code{S2} should be provided only if the user want to do the precision improvement step. In this case, \code{S2}$>$\code{S1}.

\subsubsection{Joint models for a single event}
To fit a model with only a single event, argument \code{survival\_type} should be equal to 'Single'. Arguments \code{formSurv\_02, formSurv\_12, sharedtype\_02, shardetype\_12, hazardBase\_02,} \\
\code{hazardBase\_12, delta2,  Time\_L, Time\_R} do not need to be specified. They are NULL by default.

Below are three examples of call:
\begin{Code}
lsjm1 <- lsjm(Objectlsmm = lsmmStandard, survival_type = 'Single', 
    formSurv_01 = ~ W1, sharedtype_01 = c("value", "slope") ,
    formSlopeFixed = ~ 1, formSlopeRandom = ~ 1, index_beta_slope = c(2), 
    index_b_slope = c(2), hazardBase_01 = "Splines", delta1 = ~ event, 
    Time_T = ~ Time_event, Time_T0 = ~ Time_entry, nb.knots.splines = c(3))
    
lsjm1Time2 <- lsjm(Objectlsmm = lsmmStandardTime2, survival_type = 'Single', 
    formSurv_01 = ~ W1, sharedtype_01 = c("value", "slope") ,
    formSlopeFixed = ~ I(2*time), formSlopeRandom = ~ I(2*time), 
    index_beta_slope = c(2,3), index_b_slope = c(2), hazardBase_01 = "Splines", 
    delta1 = ~ event, Time_T = ~ Time_event, Time_T0 = ~ Time_entry,
    nb.knots.splines = c(3))

lsjm2 <- lsjm(Objectlsmm = lsmmCovDep, survival_type = 'Single',
    formSurv_01 = ~ 1, sharedtype_01 = c("value", "variability"),
    hazardBase_01 = "Weibull", delta1 = ~ event, Time_T = ~ Time_event)
\end{Code}

The call \code{lsjm1} fits a joint model with a longitudinal submodel with a standard mixed model defined previously and fitted in the object \code{lsmmStandard} and a survival submodel with only one event. The hazard baseline function is defined by splines with 3 internal knots. The survival submodel is adjusted on the covariate $W1$, on the current value and current slope of the marker. This call takes into account delayed entry with an entry time defined by \code{Time\_entry}. For the slope, given that in the corresponding linear mixed model, the trajectory of the marker is defined by: 
\begin{equation}
    \widetilde{Y_{i}}(t) = \beta_0 + \beta_1 \times t + \beta_2X_{1i} + b_{0i} + b_{1i} \times t
\end{equation}
then, the slope is defined by:
\begin{equation}
    \frac{\partial \widetilde{Y_i}(t)}{\partial t}=\tilde{Y}'_i(t) = \beta_1 + b_{1i}
\end{equation}

The call \code{lsjm1Time2} fits the same joint model, but uses \code{lsmmStandardTime2} as the longitudinal submodel, thereby including \code{time}$^2$ as both a fixed and a random effect. In this case, the longitudinal trajectory is defined by:
\begin{equation}
	\widetilde{Y_i}(t) = \beta_0 + b_{0i} + (\beta_1 + b_{1i})t
	+ (\beta_2 + b_{2i})t^2 + \beta_3X_{1i},
\end{equation}
and the corresponding slope is given by:
\begin{equation}
	\frac{\partial \widetilde{Y_i}(t)}{\partial t}
	= \widetilde{Y}'_i(t)
	= \beta_1 + b_{1i} + 2(\beta_2 + b_{2i})t.
\end{equation}
Thus, in \code{lsjm1Time2}, the slope is time-dependent and is obtained by differentiating both the fixed- and random-effects components of the longitudinal trajectory with respect to time.

The call \code{lsjm2} fits a joint model combining a location-scale mixed model with covariate and time-dependent variability, defined previously and fitted in the object \code{lsmmCovDep}, and a proportional hazard submodel for only one event. The hazard baseline function is defined by a Weibull function. The survival submodel depends on the current value and variability of the marker without any other covariates.

\subsubsection{Joint models for competing events}
In the case of competing events, there are two events so \code{survival\_type} should be equal to \code{'CR'} and \code{formSurv\_02}, \code{sharedtype\_02}, \code{hazardBase\_02} and \code{delta2} must be defined. In this case, \code{Time\_T} corresponds to the date of first event (or censoring). \code{delta1} and \code{delta2} do not be both equal to 1 for one subject, only the first event is recorded.

Below are two examples of call:
\begin{Code}
lsjm3 <- lsjm(Objectlsmm = lsmmCovDep, survival_type = 'CR', 
   formSurv_01 = ~ W1, formSurv_02 = ~ W2,
   sharedtype_01 = c("value", "slope", "variability") , 
   sharedtype_02 = c("random effects", "variability") ,
   hazardBase_01 = "Splines",   hazardBase_02 = "Splines",
   formSlopeFixed = ~ 1, formSlopeRandom = ~ 1, index_beta_slope = c(2), 
   index_b_slope = c(2),  delta1 = ~ event1, delta2 = ~ event2,
   Time_T = ~ Time_event, Time_T0 = ~ Time_entry, nb.knots.splines = c(3,1))

 lsjm4 <- lsjm(Objectlsmm = lsmmInterIntra, survival_type = 'CR',
    formSurv_01 = ~ 1, formSurv_02 = ~ W1,
    sharedtype_01 = c("value", "variability inter", "variability intra"),
    sharedtype_02 = c("value", "variability inter"), 
    hazardBase_01 = "Weibull", hazardBase_02 = "Gompertz", 
    delta1 = ~ event1, delta2 = ~ event2,
    Time_T = ~ Time_event)
\end{Code}
The call \code{lsjm3} fits a joint model combining a location-scale mixed model with covariate and time-dependent variability, (defined and fitted in \code{lsmmCovDep}) and two cause-specific proportional hazard models for two competing events. Both risk functions are defined with baseline hazard functions on splines with respectively 3 and 1 internal knots for the first and the second risk. The first risk function is adjusted on covariate $W1$, on the current value, the slope and variability of the marker. The second risk function is adjusted on covariate $W2$ on the random effects and the variability of the marker. Subjects are left truncated at their time of inclusion \code{Time\_entry}.

The call \code{lsjm4} fits a joint model combining a location-scale mixed model distinguishing within and between visits residual variabilities, (defined and fitted in \code{lsmmInterIntra}) and two proportional hazard models for competing events. The first risk function is adjusted on the current value, the inter visits variability and the intra visit variability of the marker. The second risk function is adjusted on the covariate $W1$, the current value and the inter visits variability of the marker. For the baseline risk functions we consider a Weibull function for the first risk function and a Gompertz for the second one.

\subsubsection{Joint models for illness-death events}
To fit a joint model including an illness-death model, \code{survival\_type} should be equal to \code{"IDM"}, and \code{formSurv\_12, sharedtype\_12, hazardBase\_12, Time\_L} and \code{Time\_R} must be defined. In this case, \code{Time\_T} provides a one-side formula with a variable corresponding to the minimum time between time-to-transition to state $(2)$ and censoring time, \code{Time\_L} is a one-side formula with a variable giving the last observed time in state $(0)$ and \code{Time\_R} is a one-side formula with a variable providing the first time known in state $(1)$. If the subject $i$ is never seen in state $(1)$ then it is necessary, for computation reasons, to fix $R_i = T_i$. If there is no interval censoring, $L_i = R_i$ for all subjects.

Below are two examples of call:
\begin{Code}
 lsjm5 <- lsjm(Objectlsmm = lsmmCovDep, survival_type = 'IDM', 
   formSurv_01 = ~ W1, formSurv_02 = ~ W2, formSurv_12 = ~ W1,
   sharedtype_01 = c("value", "variability") , 
   sharedtype_02 = c("value","slope", "variability") ,
   sharedtype_12 = c("value") ,
   hazardBase_01 = "Splines",   hazardBase_02 = "Exponential", 
   hazardBase_12 = "Splines",
   formSlopeFixed = ~ 1, formSlopeRandom = ~ 1, index_beta_slope = c(2), 
   index_b_slope = c(2),  delta1 = ~ event1, delta2 = ~ event2,
   Time_T = ~ Time_final, Time_L = ~ Time_left, Time_R = ~ Time_right,
   Time_T0 = ~ Time_entry, nb.knots.splines = c(3,0,2))

 lsjm6 <- lsjm(Objectlsmm = lsmmInterIntra, survival_type = 'IDM', 
   formSurv_01 = ~ W1+W2, formSurv_02 = ~ 1, formSurv_12 = ~ W3,
   sharedtype_01 = c("value", "variability inter") , 
   sharedtype_02 = c("value","slope", "variability intra") ,
   sharedtype_12 = c("value", "variability inter", "variability intra") ,
   hazardBase_01 = "Weibull",   hazardBase_02 = "Exponential", 
   hazardBase_12 = "Gompertz",
   formSlopeFixed = ~ 1, formSlopeRandom = ~ 1, index_beta_slope = c(2), 
   index_b_slope = c(2),  delta1 = ~ event1, delta2 = ~ event2,
   Time_T = ~ Time_final, Time_L = ~ Time_left, Time_R = ~ Time_right)
\end{Code}
The call \code{lsjm5} fits a joint model combining a location-scale mixed model with covariate and time-dependent variability, (defined and fitted in \code{lsmmCovDep}) and an illness-death model with 3 transitions. For transitions $(01)$ and $(12)$, the baseline hazard functions are defined on splines with respectively 3 and 2 internal knots. These two transitions intensities are adjusted on the same covariate $W1$ and on the current value of the marker. Transition $(01)$ is also adjusted on the current residual variability. Transition $(02)$ is defined using an exponential baseline hazard function and is adjusted on covariate $W2$, the current value, the slope and the current residual variability of the marker. Subjects are left truncated (\code{Time\_T0 = ~Time\_entry}). The last visit seeing healthy is defined by covariate \code{Time\_left} and the first visit seeing in state $(1)$ by \code{Time\_right}. The last time of information is \code{Time\_final}.\\

The call \code{lsjm6} fits a joint model combining a location-scale mixed model with between-visits and within-visit residual variabilities, defined and fitted in \code{lsmmInterIntra} and an illness-death model with 3 transitions. The baseline hazard functions are respectively a Weibull, an Exponential and a Gompertz function. Transition $(01)$ is adjusted on \code{W1}, \code{W2}, the current value and the between-visits variability of the marker. Transition $(02)$ is adjusted on the current value, the slope and the between-visits variability of the marker. Finally, transition $(12)$ is adjusted on \code{W3}, the current value, and on both between-visits and within-visit variabilities of the marker.

\subsection{Random effects prediction: generic function ranef}
Function \fct{ranef} returns the predicted random effects for all subjects.\\

Below is an example of call:
\begin{Code}
re <- ranef(m1)
\end{Code}
The \code{m1} object is either a \code{lsmm} object or a \code{lsjm} object. 

\subsection{Longitudinal and survival predictions: generic function predict}
The \code{predict} function provides several predictions. Its syntax is
\begin{Code}
predict(object, which = c('RE','Y','Cum'), Objectranef = NULL, data.long = NULL)
\end{Code}
where:
\begin{itemize}
    \item[$\bullet$] \code{object} provides either a \code{lsmm} object or a \code{lsjm} object,
    \item[$\bullet$] \code{which} indicates which predictions are computed. \code{'RE'} corresponds to the random effects (equivalent to use \fct{ranef} function), \code{'Y'} to the expected marker and \code{'Cum'} to the cumulative risk functions,
    \item[$\bullet$] \code{data.long} contains the longitudinal dataset used for making predictions. If \code{data.long} is not provided, the predictions are performed on the data used to estimate the model.
\end{itemize}
First, with the option \code{which = "RE"} it provides the prediction of the random effects as the \code{ranef} function.
Secondly, with the option \code{which = "Y"} it computes the prediction of the marker $Y$. This function returns individual predicted values of $Y$ at each subject's observed visit times, which may differ across subjects. The option \code{which = c("RE", "Y")} indicates to compute both the predictions of random effects and of the marker. In this case, the function will returns two tables, one for the random effects and one for the marker.
Finally, with the option \code{which = "Cum"} it computes the predicted cumulative risk function(s). The function returns as many tables as transitions with in each table the predicted value for the cumulative risk function corresponding to the transition.

Below are four examples of call:
\begin{Code}
predY.1 <- predict(m1, which = c("RE", "Y"))
predY.2 <- predict(m1, which = c("Y"), ranefObject = re)
predCum <- predict(m1, which = c("Cum"), ranefObject = re)
predTot <- predict(m2, which = c("RE", "Y","Cum"))
\end{Code}
The \code{m1} object is either a \code{lsmm} object or a \code{lsjm} object, whereas the \code{m2} object must be a \code{lsjm} object. The call \code{predY.1} returns two tables, one for the predicted random effects and one for the predicted values of the marker. The calls \code{predY.2} and \code{predCum} use the previously predicted random effects \code{re} which is an object of the \code{ranef} function. The call \code{predTot} computes all possible predictions.

\subsection{Goodness-of-fit: generic function plot}
The \fct{plot} function provides several graphics. The syntax of the function is 
\begin{Code}
plot(x, which = c('long.fit', 'survival.fit', 'traj.ind'), Objectpredict, 
             break.times = NULL, ID.ind = NULL, ObjectSmoothHazard = NULL, ...)
\end{Code}
with the following arguments:
\begin{itemize}
    \item[$\bullet$] \code{x} either a \code{lsmm} object or a \code{lsjm} object,
    \item[$\bullet$] \code{which} a character indicating which plot must be displayed,
    \item[$\bullet$] \code{Objectpredict} an object of the \code{predict} function,
    \item[$\bullet$] \code{break.times} a vector of breaking times to create windows if \code{which = 'long.fit'},
    \item[$\bullet$] \code{ID.ind} a vector providing the id of subjects for whom we wish to trace the individual trajectory.
    \item[$\bullet$] \code{ObjectSmoothHazard} A SmoothHazard object (only for IDM survival model)
\end{itemize}

First, if argument \code{which = "long.fit"} \code{plot} function allows to assess the fit of the longitudinal submodel comparing the mean of marker predictions collected in some windows of times (defined by \code{break.times} or using percentiles) to the mean of the observed measurements and its 95\% confidence interval.

Then, for a \code{lsjm} object only, if argument \code{which = "survival.fit"}, \code{plot} function allows to assess the fit of the survival submodel. For each transition, the predicted cumulative hazard function is computed for each subject at each event time given the predicted random effects. Then the mean of the predicted cumulative hazard functions are compared with there Nelson-Aalen estimator in the case of a single event or with competing risks. For an illness-death model, the predicted cumulative hazard function for each transition is compared to an illness-death model estimated by penalized likelihood accounting for interval censoring with the \pkg{SmoothHazard} package \citep{touraine_2017}.

Lastly, with the option \code{which = "traj.ind"} the \fct{plot} function represents the individual trajectory with its prediction interval.

Below are three examples of call:
\begin{Code}
plot(m1, which = 'long.fit', Objectpredict = predY.2, break.times = vec_time)

plot(m2, which = 'survival.fit', Objectpredict = predTot, 
                                     ObjectSmoothHazard = fit.smooth)

plot(m1, which = 'traj.ind', Objectpredict = predY.2, ID.ind = vec_ind)
\end{Code}

The \code{m1} object is either a \code{lsmm} object or a \code{lsjm} object, whereas the \code{m2} object must be a \code{lsjm} object. Argument \code{Objectpredict} provides the predicted random effects, marker values and cumulative hazard functions obtained with \fct{predict} function. The argument \code{break.times} provides the vector of breaking times to create windows. If this argument is not provided, the windows are created taking the percentiles of the visit times. The argument \code{ObjectSmoothHazard} provides the estimation of the illness-death model obtained with \pkg{SmoothHazard} package. The argument \code{ID.ind} is a vector providing the id of subjects for whom we wish to trace the individual trajectory. 

\subsection{Dynamic prediction of the event: function dynpred}

The \fct{dynpred} function allows to compute individual dynamic prediction of an event for any subject, included or not in the dataset used for estimating the model. The syntax of \fct{dynpred} is 
\begin{Code}
dynpred(object, newdata,  s, horizon, event, CI = 95, nb.draws = 1000)
\end{Code}
with the following arguments:
\begin{itemize}
	\item[$\bullet$] \code{object} an \code{lsjm} object.
    \item[$\bullet$] \code{newdata} a dataset containing the individuals for whom predictions are expected.
    \item[$\bullet$] \code{s} the landmark time (a single value).
    \item[$\bullet$] \code{horizon} the horizon time of prediction: the probability of having the event between \code{s} and \code{horizon} is computed. It could be a vector or a single value.
    \item[$\bullet$] \code{event} an integer indicating for which event the prediction is computed. In the case of a competing risk model, it could be equal to 1 or 2 whereas for both other models it could be equal to 1 only.
    \item[$\bullet$] \code{CI} an integer between 0 and 100 indicating the level for computing the confidence interval. By default, \code{CI = 95} for a 95\% confidence interval. If \code{CI = NULL}, then the confidence interval is not computed.
    \item[$\bullet$] \code{nb.draws} an integer providing the number of draws used to compute the confidence interval, 1000 by default.
\end{itemize}

Below is an example of call:
\begin{Code}
predyn <- dynpred( m2, data2, s = 3, horizon = c(4,5), event = 1, 
    IC = 95, nb.draws = 1000)
\end{Code}
In this example, \code{m2} is a \fct{lsjm} model. The predictions are computed for all subjects in \code{data2} for the first event between times 3 and 4 and between time 3 and 5. The 95\% confidence interval is computed with 1000 draws.

\subsection{Population-average probability of being free of any event: function survmarg}

The \fct{survmarg} function allows to compute the average probability of being free of any events at a given time $t$ and given covariates. The syntax of \fct{survmarg} is 
\begin{Code}
	survmarg(object, individual,  time)
\end{Code}
with the following arguments:
\begin{itemize}
	\item[$\bullet$] \code{object} an \code{lsjm} object.
	\item[$\bullet$] \code{individual} a subject profile for which the probability is computed
	\item[$\bullet$] \code{time} a numeric, the time $t$ at which the probability is computed
\end{itemize}

Below is an example of call:
\begin{Code}
	profile <- data[which(data$ID == 1),]
	survM <- survmarg(m2, profile, time = 3)
\end{Code}

In this example, \code{m2} is a \fct{lsjm} model. The probability of being free of any events at time 3 is computed for the specific-profile defined by \code{profile}, which corresponds to the covariate values of individual 1.

\section{Application on real data}
This section presents some examples of the application of \pkg{LSJM} \proglang{R}-package functions, whether for model estimations, graphical outputs, or dynamic predictions. These examples are illustrated using the real \code{threeC} data supplied with our \proglang{R}-package.\\

After installing the package, the first step consists in loading package \pkg{LSJM}:
\begin{Code}
library("LSJM")
\end{Code}
\subsection{ThreeC data}
The \code{threeC} dataset from \pkg{LSJM} \proglang{R}-package is a random subsample of 500 subjects, identified by \code{ID}, from the French Three-Cities cohort aimed at assessing the relation between vascular factors and dementia in the elderly \citep{3c_2003}. The sample include participants without dementia at baseline and aged 65 years old or older. Repeated measures of systolic blood pressure (\code{SBP}) were collected over a maximum period of 20 years. At each visit (visit ID represented by the variable \code{num.visit}), systolic blood pressure was measured two or three times. The age at entry in the cohort (\code{age0}), the age at the visit (\code{age.visit}), the age of last visit without dementia (\code{age.last}), the age of visit diagnosis  (\code{age.first}), the age of death or censoring (\code{age.final}) and the indicator of dementia and death (\code{dem}, \code{death}) were also collected. The sex variable (\code{sex}) is also reported, with 187 men and 313 women. Educational level (\code{Edu}) is categorized according to the number of years of education, with 201 participants having more than 10 years of study and 299 having 10 years or less. Finally, the APOE4 genotype (\code{Apoe4}) is reported, with 87 participants carrying the APOE4 allele and 413 not carrying it.\\
For computation and interpretation purposes, all \code{age} variables will be replaced by the age minus 65 divided by 10. The values of blood pressure are also divided by 10 which leads to a measurement unit in cmHg instead of mmHg.\\
The following lines create the variables relative to the new \code{age}, and display the first lines of the \code{threeC} dataset:\\

\begin{Code}
data(threeC)
threeC$age.visit65 <- (threeC$age.visit-65)/10
threeC$age.final65 <- (threeC$age.final-65)/10
threeC$age0_65 <- (threeC$age0-65)/10
threeC$age.last65 <- (threeC$age.last-65)/10
threeC$age.first65 <- (threeC$age.first-65)/10
threeC$SBP <- threeC$SBP/10

head(threeC)

ID age.visit num.visit  SBP age.final     age0 age.last age.first dem death   sex 
1  72.17522        S0 14.2  90.44216 72.17522 89.85352  89.85352   0     1   Woman   
1  74.57632        S1 13.4  90.44216 72.17522 89.85352  89.85352   0     1   Woman   
1  74.57632        S1 12.8  90.44216 72.17522 89.85352  89.85352   0     1   Woman   
1  76.56400        S2 14.2  90.44216 72.17522 89.85352  89.85352   0     1   Woman  
1  76.56400        S2 12.0  90.44216 72.17522 89.85352  89.85352   0     1   Woman  
1  76.56400        S2 11.5  90.44216 72.17522 89.85352  89.85352   0     1   Woman   
      Edu Apoe4 age.visit65 age.final65   age0_65 age.last65 age.first65
	0     0   0.7175222    2.544216 0.7175222   2.485352    2.485352
	0     0   0.9576318    2.544216 0.7175222   2.485352    2.485352
	0     0   0.9576318    2.544216 0.7175222   2.485352    2.485352
	0     0   1.1563997    2.544216 0.7175222   2.485352    2.485352
	0     0   1.1563997    2.544216 0.7175222   2.485352    2.485352
	0     0   1.1563997    2.544216 0.7175222   2.485352    2.485352
\end{Code}

\subsection{A LSJM with time-dependent variability and an illness-death model}
This subsection illustrates the estimation of a LSJM. The LSMM sub-model is estimated suggesting a covariate and time-dependent variability and the survival sub-model is an illness-death model.

\subsubsection{Data management}
In this example, we consider only one measure at each visit. So we create first a new variable: the mean of all measures collected at the corresponding visit, \code{SBPmean}. Then, we delete the measurements of blood pressure collected after the diagnosis time of dementia. The following lines create the new dataset and display its first lines:

\begin{Code}
library(dplyr)
threeC_ex1 <- threeC
threeC_ex1 <- threeC_ex1 %>% group_by(ID, num.visit) %>% mutate(SBPvisit =
                                                            mean(SBP))
threeC_ex1 <- threeC_ex1[!duplicated(threeC_ex1[, c("ID", "num.visit")]),
                         c("ID", "SBPvisit", "age.visit65", "age0_65", 
                         "age.final65","age.last65",  "age.first65", "dem", 
                         "death", "sex")]
                           
head(threeC_ex1)                           

ID  SBPvisit age.visit65 age0_65 age.final65 age.last65 age.first65   
1     14.2       0.718   0.718        2.54       2.49        2.49     
1     13.1       0.958   0.718        2.54       2.49        2.49     
1     12.6       1.16    0.718        2.54       2.49        2.49     
1     11.7       1.48    0.718        2.54       2.49        2.49     
1     12.6       1.73    0.718        2.54       2.49        2.49     
1     10.8       2.00    0.718        2.54       2.49        2.49    
1     12.4       2.19    0.718        2.54       2.49        2.49     
1     12.7       2.49    0.718        2.54       2.49        2.49     
2     14         0.975   0.975        2.07       1.21        2.02    
2     12.3       1.21    0.975        2.07       1.21        2.02     
 dem death sex
  0     1  woman 
  0     1  woman 
  0     1  woman 
  0     1  woman 
  0     1  woman 
  0     1  woman 
  0     1  woman 
  0     1  woman 
  1     1  woman 
  1     1  woman 
\end{Code}
Here, subject 1 is a woman, entered in the study at 0.718, who was not diagnosed with dementia and died at time 2.54, whereas individual 2 was diagnosed with dementia at 2.02 and died at 2.07.\\ 

\subsubsection{LSMM model}
\label{lsmm1}
The first step is to fit a LSMM using the \fct{lsmm} function with linear trajectories of \code{SBP} and its residual variability with \code{age.visit}, assuming uncorrelated random effects between the random effects of the mean and the one of the variance. The next lines estimates the corresponding model:
\begin{Code}
m1 <- lsmm(formFixed = SBPvisit ~ age.visit65,
              formRandom = ~ age.visit65,
              formGroup = ~ ID,
              timeVar = 'age.visit65',
              data.long = threeC_ex1,
              formVar = "cov-dependent",
              formFixedVar = ~ age.visit65,
              formRandomVar = ~ age.visit65,
              correlated_re = FALSE,
              S1 = 500,
              S2 = 1000,
              nproc = 2)

summary(m1)
Location-scale linear mixed model fitted by maximum likelihood method 

Statistical Model: 
    Number of subjects: 500 
    Number of observations: 2458 

Iteration process: 
    Convergence criteria satisfied
     Number of iterations:  
          Step 1:  17 
          Step 2:  1 
     Convergence criteria (Step1): parameters = 4.93e-09 
                                 : likelihood = 2.92e-07 
                                 : second derivatives = 6.4e-14 
     Time of computation : 1.145953 mins 

Goodness-of-fit statistics:
    Likelihood:  -5027.3 
    AIC:  10074.599 

Maximum Likelihood Estimates:
Longitudinal model:
      Fixed effects of the location part:
              Coeff     SE     Wald Pvalue
(Intercept) 13.8661 0.1387 100.0068 <0.001
age.visit65  0.2868 0.0836   3.4324 <0.001

     Fixed effects of the scale part:
             Coeff     SE   Wald Pvalue
(Intercept) 0.2262 0.0555 4.0739 <0.001
age.visit65 0.1183 0.0331 3.5699 <0.001

     Covariance matrix of the location random effects:
            (Intercept) age.visit65
(Intercept)    4.283720   -1.757586
age.visit65   -1.757586    1.157830

     Covariance matrix of the scale random effects:
            (Intercept)  age.visit65
(Intercept)  0.08713937 -0.016774493
age.visit65 -0.01677449  0.007028832 
\end{Code}
First, the \fct{summary} method details the dataset, the number of subjects and the number of observations. The next block provides information about the convergence process with the number of iterations for both step 1 and 2, the convergence criteria of step 1, the time of computation and whether the model converged correctly ("Convergence criteria satisfied"). Then some goodness-of-fit statistics are given: the maximum log-likelihood and the Akaike information criteria (AIC). Finally, tables of estimates are provided including the estimated parameter (\code{Coeff}), the estimated standard error (\code{SE}), the Wald test statistics with the normal approximation (\code{Wald}) and the corresponding p-value (\code{Pvalue}), for both the fixed effects associated with the mean (location part) and the fixed effects associated with the residual variability (scale part). For the random effect distributions, we distinguish two estimated covariance matrices, one for the location part and one for the scale part of the model, since these two parts have been assumed to be uncorrelated (\code{correlated\_re = FALSE}). If \code{correlated\_re = TRUE}, a single covariance matrix is returned for all the random effects in the model.\\

All estimated LSMM parameters and their standard errors are available in the \code{table.res} object of the list returned by the \fct{lsmm} function. This concerns the fixed parameters, the parameters of the Cholesky decomposition of the random effects covariance matrix, as well as the parameters of the latter obtained via the Delta-method. The  prediction of the random effects and the predicted value of the marker for each individual at their respective measurement times are computed using the \fct{ranef} or \fct{predict} functions and the trajectory of the marker can be displayed by the \fct{plot} function. Since the same functions can be applied to LSJM models, we will detail their usage in the following section.

\subsubsection{LSJM model}
Then \fct{lsjm} function allows to estimate the illness-death joint model, assuming baseline hazard functions defined using a Weibull function for transitions $(01)$ and $(02)$ and splines with one internal knot for transition $(12)$. Data are left truncated by the age of entry in the study, \code{age0\_65}. All transitions are adjusted on the current value and the variability of the marker. The transition from healthy to death is also adjusted on the slope. And finally, both transitions to death are adjusted on the sex. The next lines estimate the corresponding model:

\begin{Code}
l1 <- lsjm(m1,
           survival_type = 'IDM',
           formSurv_01=~1,
           formSurv_02=~sex,
           formSurv_12=~sex,
           sharedtype_01 = c("value", "variability"),
           sharedtype_02 = c("value", "slope", "variability"),
           sharedtype_12 = c("value", "variability"),
           hazardBase_01 = "Weibull",
           hazardBase_02 = "Weibull",
           hazardBase_12 = "Splines",
           delta1=~dem,
           delta2=~death,
           Time_T =~age.final65,
           Time_L =~age.last65,
           Time_R =~age.first65,
           Time_T0 =~age0_65,
           formSlopeFixed =~1,
           formSlopeRandom =~1,
           index_beta_slope = c(2),
           index_b_slope = c(2),
           nb.knots.splines = c(0,0,1),
           S1 = 1000,
           S2 = 2000,
           nproc = 10)

summary(l1)
Location-scale joint model for an illness-death model
fitted by maximum likelihood method 

Statistical Model: 
    Number of subjects:  500
    Number of observations:  2458

Iteration process: 
    Convergence criteria satisfied
     Number of iterations:  
          Step 1:  20 
          Step 2:  1 
     Convergence criteria (Step1): parameters = 4.47e-05 
                                 : likelihood = 9.44e-06 
                                 : second derivatives = 1.76e-10 
     Time of computation : 6.874512 hours 

Goodness-of-fit statistics:
    Likelihood:  -5883.375 
    AIC:  11822.751 

Maximum Likelihood Estimates:
Longitudinal model:
      Fixed effects of the location part:
              Coeff     SE    Wald Pvalue
(Intercept) 13.9551 0.1566 89.1354 <0.001
age.visit65  0.2131 0.1083  1.9680  0.049

     Fixed effects of the scale part:
             Coeff     SE   Wald Pvalue
(Intercept) 0.1537 0.0538 2.8568  0.004
age.visit65 0.1791 0.0339 5.2860 <0.001

     Covariance matrix of the location random effects:
            (Intercept) age.visit65
(Intercept)    4.366972   -1.810359
age.visit65   -1.810359    1.193415

     Covariance matrix of the scale random effects:
             (Intercept)   age.visit65
(Intercept)  0.065553512 -0.0037253869
age.visit65 -0.003725387  0.0004496534

Survival models:
    Transition 0-1:
       Regression:
                 Coeff     SE    Wald Pvalue
value 01       -0.0342 0.0933 -0.3669  0.714
variability 01  1.6529 0.7477  2.2108  0.027

     Baseline:  Weibull 

            Coeff     SE    Wald Pvalue
intercept -6.7827 1.3252 -5.1184 <0.001
shape_01   2.0655 0.0954 21.6561 <0.001

    Transition 0-2:
       Regression:
                 Coeff     SE    Wald Pvalue
value 02       -0.3480 0.1459 -2.3846  0.017
slope 02       -0.0877 0.2257 -0.3885  0.698
variability 02  2.1693 0.9036  2.4007  0.016
sexWoman_02    -0.9835 0.2240 -4.3909 <0.001

     Baseline:  Weibull 

            Coeff     SE    Wald Pvalue
intercept -2.0109 2.0339 -0.9887  0.323
shape_02   1.8216 0.0992 18.3585 <0.001


    Transition 1-2:
       Regression:
                 Coeff     SE    Wald Pvalue
value 12        0.1028 0.0719  1.4301  0.153
variability 12 -0.1229 0.4026 -0.3052   0.76
sexWoman_12    -0.7427 0.1907 -3.8939 <0.001

     Baseline:  Splines 

              Coeff     SE    Wald Pvalue
splines12_1 -6.2992 7.5075 -0.8390  0.401
splines12_2 -0.7688 2.7564 -0.2789  0.780
splines12_3 -0.3364 2.3142 -0.1454  0.884
splines12_4  0.5493 1.7673  0.3108  0.756
splines12_5  4.1182 4.1341  0.9961  0.319
\end{Code}
Similarly to the LSMM, the first part of the \fct{summary} method provides details about the dataset, the convergence process and the results of the longitudinal submodel. For this example, the convergence criteria where satisfied in 20 iterations for the first step and 1 for the second. There are 500 subjects and 2458 obervations. The AIC is equal to 11822.751. According to the table on the fixed effects of the scale part, blood pressure residual variability increases with age. The last part of the summary gives the estimated values of the parameters associated with each transition functions: the regression parameters and the baseline parameters. First, for the transition to dementia, it appears that dementia occurrence seems to be not associated with the current value of blood pressure (Estimate = -0.0342 and Pvalue = 0.714) but the risk increases with the variability of blood pressure (Estimate = 1.653 and Pvalue = 0.027). Then, the risk of death without dementia decreases with the current value of blood pressure (Estimate = -0.348 and Pvalue = 0.017) but increases with the variability (Estimate = 2.169 and Pvalue = 0.016). This risk is also lower for women (Estimate = -0.984 and Pvalue $<$0.001). Finally, the risk of death with dementia seems to be associated only with the sex (Estimate = -0.743 and Pvalue $<$0.001). The estimated Cholesky parameters, their standard errors and the standard errors of the covariance matrices obtained with the Delta-Method are given by \code{l1\$table.res}.\\

Then, \code{predict} function allows to compute the empirical Bayes estimates of the random effects, the predicted values for each individual at their respective visiting times and the cumulative risks for each considered transitions:

\begin{Code}
predictl1 <- predict(l1, which = c('RE','Y','Cum'))

head(predictl1$predictRE)
id (Intercept)_Location age.visit65_Location (Intercept)_Scale age.visit65_Scale
 2           -0.7218100            0.1361275       0.05100922     -0.002302383
 5            1.3343421            0.2648115       0.22039363     -0.010803398
13            1.0049743           -0.4892723       0.20685208     -0.010553859
18            1.7085881           -0.8189352      -0.16157812      0.006838082
21            0.2941827           -0.8438771      -0.02399378      0.001073053
32           -1.9701846            0.4015402       0.03410410     -0.001663643

head(predictl1$predictY)
id      time    predY   predSD
 2 0.9754278 13.57392 1.458094
 2 1.2138946 13.65720 1.520873
 5 1.5854209 16.04713 1.898126
 5 1.8156742 16.15717 1.973113
 5 2.0160849 16.25296 2.040788
13 0.4451061 14.83712 1.545819
\end{Code}

\begin{figure}[!ht]
    \centering
    \includegraphics[scale = 0.4]{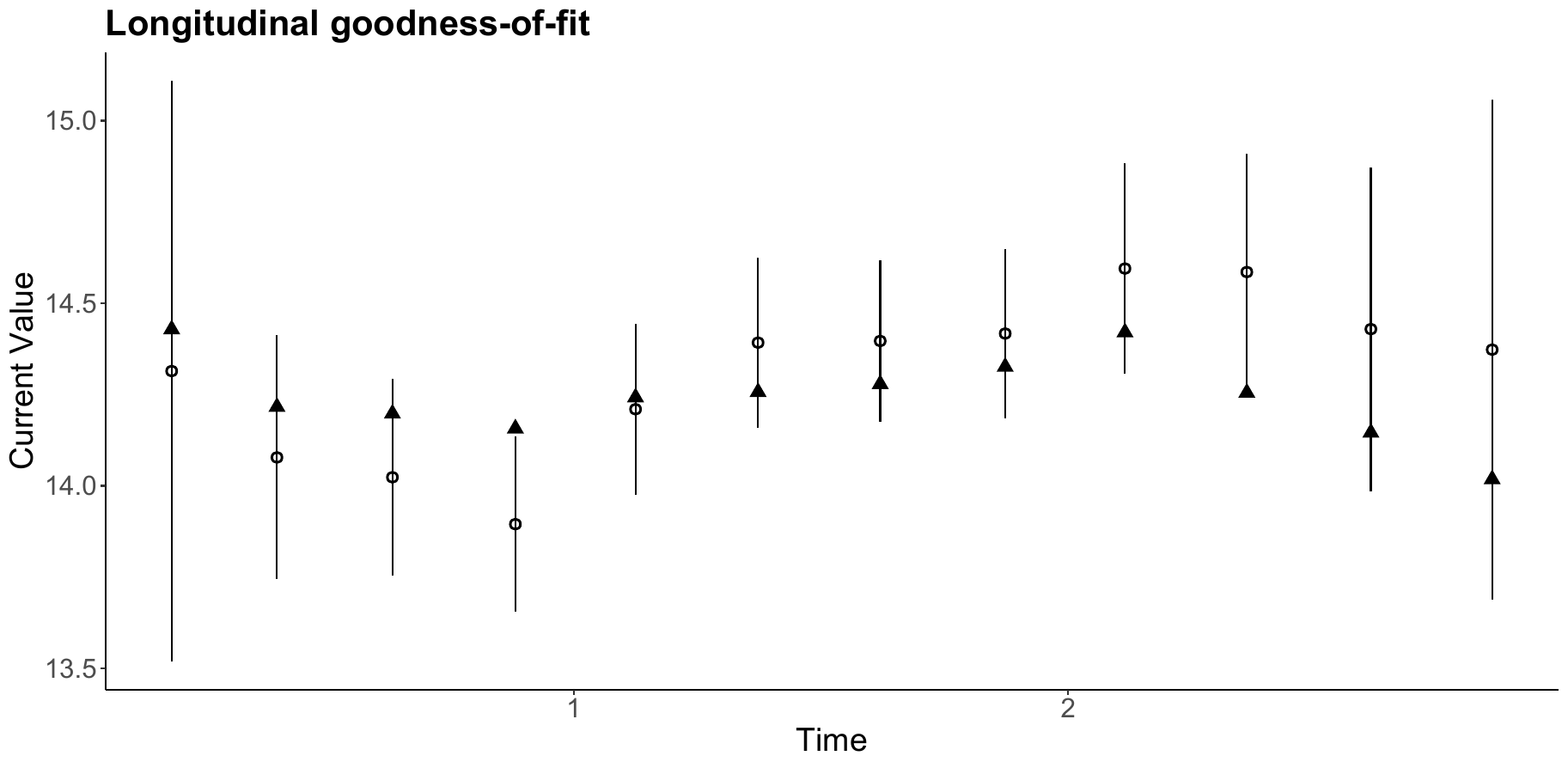}
    \caption{Comparison between predicted values of the marker from the joint model (black triangles) and the observations (mean in white circles with 95\% confidence interval (by 2.5-year age intervals).}
    \label{LongfitEx1}
\end{figure}

\begin{figure}[!ht]
	\centering
	
	\begin{subfigure}{0.48\textwidth}
		\centering
		\includegraphics[width=\textwidth]{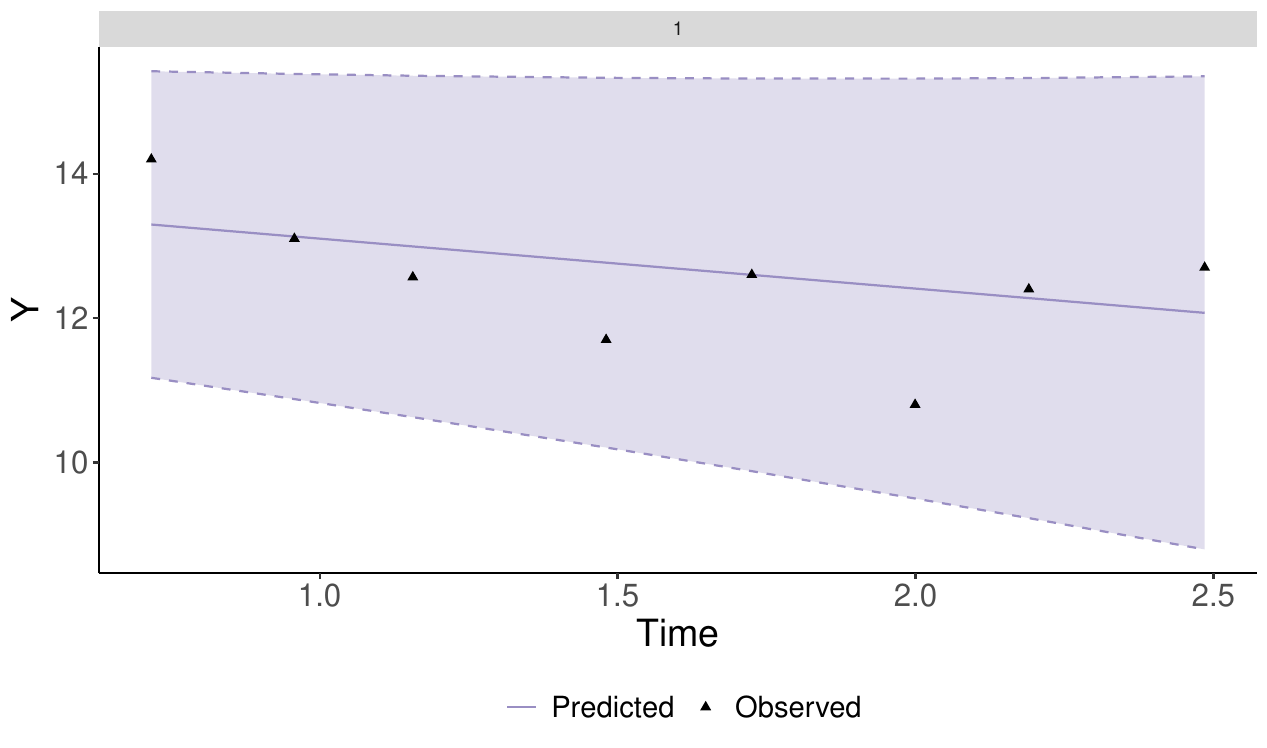}
	\end{subfigure}
	\hfill
	\begin{subfigure}{0.48\textwidth}
		\centering
		\includegraphics[width=\textwidth]{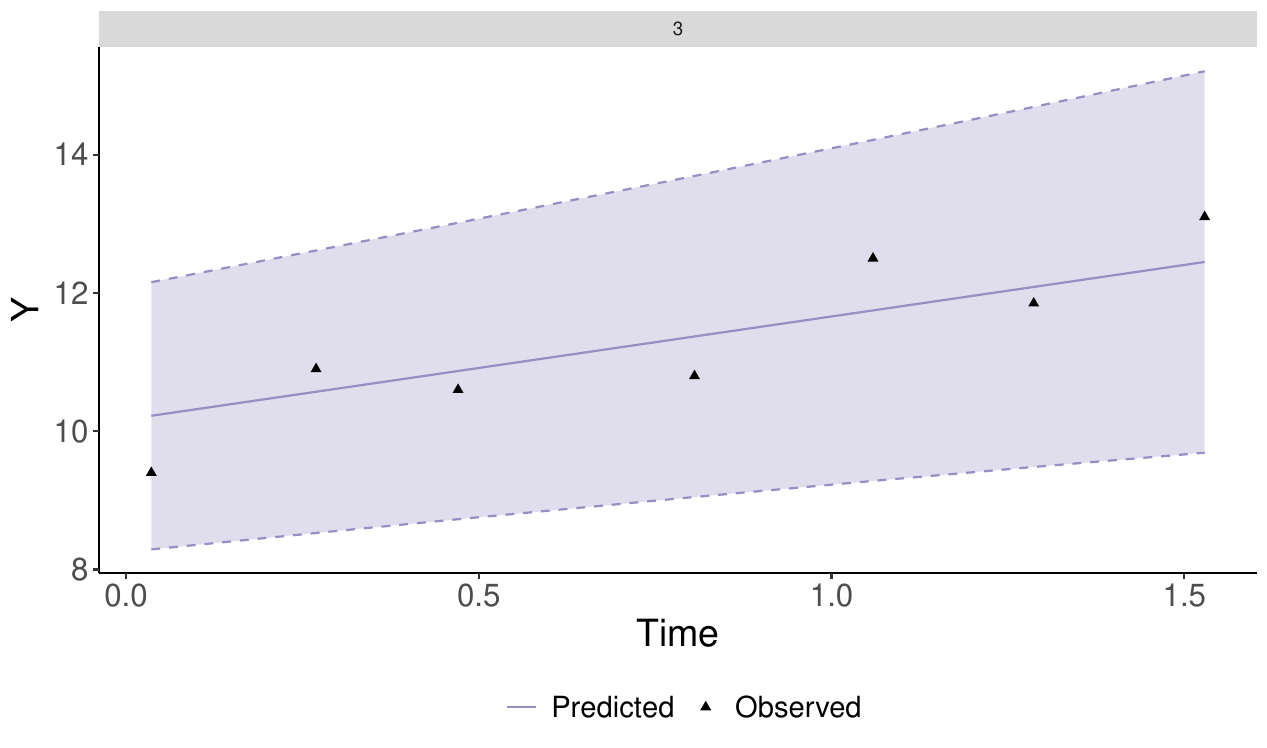}
	\end{subfigure}
	
	\caption{Prediction over time of the individual blood pressure and its prediction interval at 95\% for two subjects. The black triangles are the observed measurements.}
	\label{Ind_IDM}
\end{figure}

The goodness-of-fit of the longitudinal part can be assess by displaying the mean of marker predictions at each visit times and the mean of the observed measurements as in Figure \ref{LongfitEx1}. Then, Figure \ref{Ind_IDM} presents the predicted trajectory and its confidence interval for two selected individual. The following lines provide the code to create these graphics:
\begin{Code}

plot(l1, which = "long.fit", Objectpredict = predictl1,  
                                 break.times = (seq(65,95,by = 2.5)-65)/10)
plot(l1, which = "traj.ind", Objectpredict = predictl1,  
                                                  ID.ind = c(1,3))

\end{Code}

\begin{figure}[!ht]
    \centering
    \includegraphics[scale = 0.4]{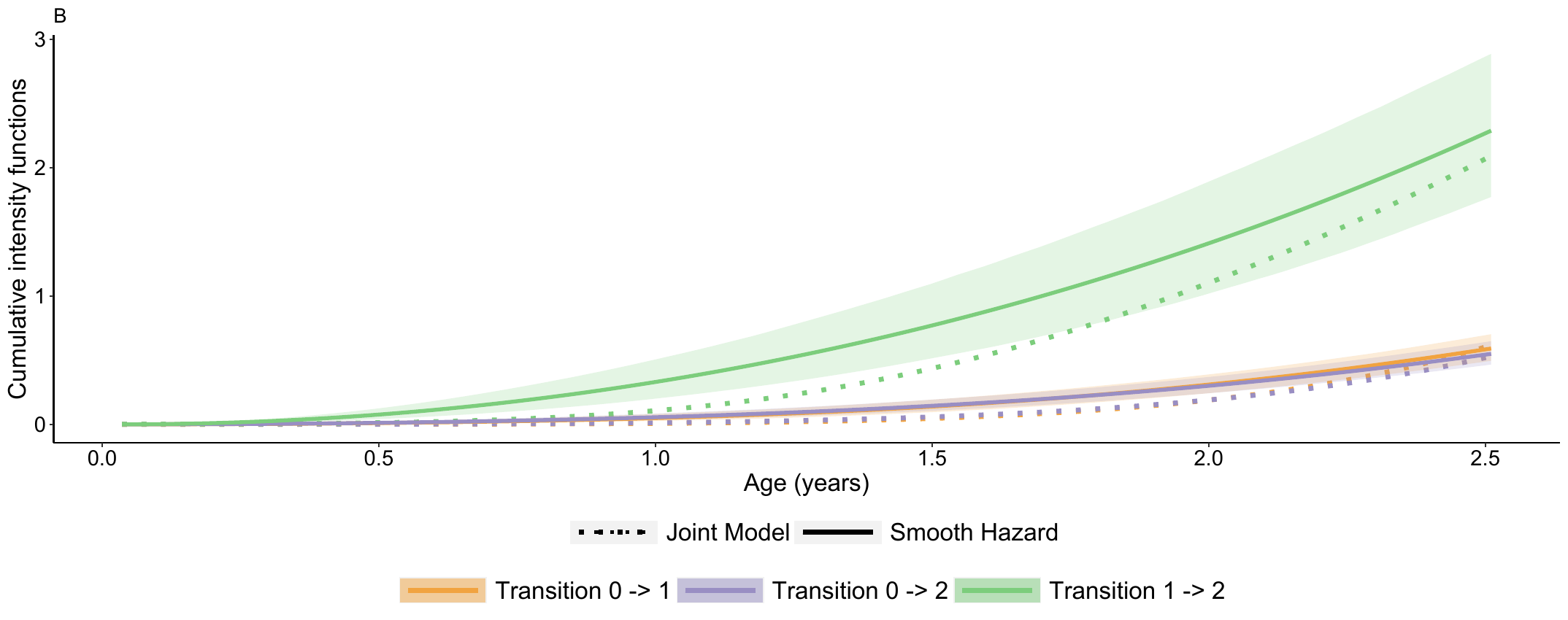}
    \caption{Comparison between the predicted cumulative hazard function from the joint model for each transition and a non-parametric estimator with 95\% confidence interval (Illness-death model estimated by penalized likelihood accounting for interval censoring with the \proglang{R} package \pkg{SmoothHazard}).}
    \label{Surv_IDM}
\end{figure}

For the survival part, the mean of the predicted cumulative hazard function
should be compared for each transition to a non-parametric estimator obtained using the \pkg{SmoothHazard} package \citep{touraine_2017}. The following lines provide the code to create the associated graphic (presented in Figure \ref{Surv_IDM}). This figure show that the adjustment between the model and the data could be improved.
\begin{Code}
library(SmoothHazard)
threeC_ex1.id <- threeC_ex1[!duplicated(threeC_ex1$ID),]

sm1 <- idm(formula02=Hist(time=age.final65,event=death,entry=age0_65)~1,
           formula01=Hist(time=list(age.last65,age.first65),event=dem, 
           entry = age0_65)~1,
           formula12=Hist(time=age.final65,event=death,entry=age0_65)~1,
           method="Splines", CV = 1, n.knots = c(15,15,15), 
           kappa = c(10, 10, 2), data=threeC_ex1.id)
                       
plot(l1, which = 'survival.fit', Objectpredict = predictl1,
                                           ObjectSmoothHazard = sm1)

\end{Code}

\begin{figure}[!ht]
    \centering
    \includegraphics[scale = 0.5]{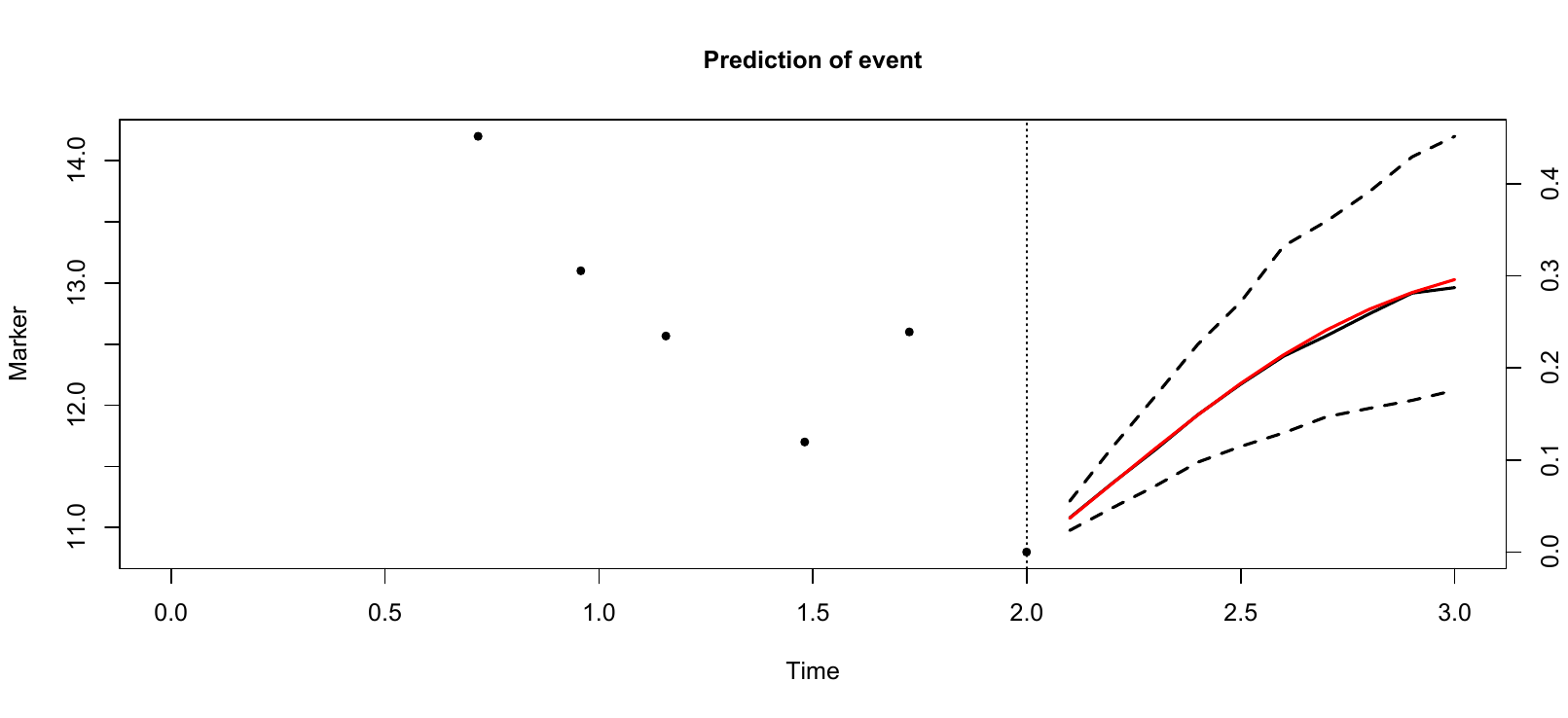}
    \caption{Prediction of the risks of Dementia between 2 and 3 years (with its 95\% confidence interval indicated by dashed lines), for a patient at risk at 2 years.}
    \label{Dynpred_IDM}
\end{figure}

Finally, the \fct{dynpred} function allows to compute the individual dynamic prediction of an event. As described in section \ref{DynpredSection}, for a specific subject (e.g a new one), whose data are provided in the input, the probabilities of occurrence of the first event from a landmark time indicated in \code{s} at horizons indicated in \code{horizon} are computed using the estimated model and the longitudinal information of the new subject up to the landmark time \code{s}. The following example is for a subject of the estimation dataset but it could be perform for a new subject. We compute the probability that dementia occurs between landmark time 2 and time 3. The prediction is computed for each time point separated by 0.1 time unit between 2 and 3, allowing us to obtain a smooth prediction curve. We also compute the confidence interval of this prediction. Figure \ref{Dynpred_IDM} illustrates the result, and the following lines provide the code for one individual:

\begin{Code}
ind1 <- as.data.frame(threeC_ex1[which(threeC_ex1$ID == 1),])
dynp <- dynpred(l1, ind1,  s = 2, horizon = seq(2.1,3,0.1),
        nb.draws = 1000)

dynp
ID Time Prediction     Median      ICinf      ICsup Empirical SD
1  2.1 0.03668934 0.03711915 0.02341961 0.05515911  0.008347391
1  2.2 0.07454292 0.07489770 0.04803115 0.11390105  0.016467563
1  2.3 0.11238074 0.11093355 0.07166993 0.16920983  0.025350245
1  2.4 0.14898390 0.14924575 0.09761294 0.22580855  0.033167628
1  2.5 0.18321395 0.18236693 0.11472371 0.27167955  0.040913491
1  2.6 0.21412335 0.21263945 0.12935679 0.33223693  0.049535750
1  2.7 0.24103969 0.23474732 0.14684326 0.35918877  0.054841381
1  2.8 0.26361194 0.25857337 0.15603418 0.39115197  0.061001419
1  2.9 0.28181406 0.28113784 0.16466651 0.42909889  0.067716372
1  3.0 0.29590811 0.28717605 0.17564340 0.45149581  0.070177043
\end{Code}

\subsection{A LSJM model distinguishing inter and intra visit variabilities for competing events}

This second example illustrates the estimation of a LSJM combining within and between visits variabilities and competing events by studying blood pressure trajectory and the risk of dementia and death. 

\subsubsection{Data management}
As we consider dementia and death as competing risk without handling interval censoring, we suppose as known the exact age of dementia onset taking the middle of interval between the last visit without dementia and the visit of diagnosis \footnote{this corresponds to the naive method discussed in section \ref{init} to present the parameter initialization of the illness-death model dealing with interval censoring;}. If an individual is diagnosed with dementia, he is then censored for death and conversely. Then, as the example distinguishes within and between visits variabilities, all measurements are kept, excepted those collected after the time of the event. The following lines create the new dataset and display its first lines:\\

\begin{Code}
threeC_ex2 <- threeC[,c("ID", "SBP", "age.visit65", "age0_65",
                        "age.final65", "age.last65", "age.first65", 
                        "dem", "death", "sex","num.visit")]

threeC_ex2$age65_CR <- NA
threeC_ex2$age65_CR[which(threeC_ex2$dem == 1)] <-
  (threeC_ex2$age.last65[which(threeC_ex2$dem == 1)] +
     threeC_ex2$age.first65[which(threeC_ex2$dem == 1)])/2
threeC_ex2$age65_CR[which(threeC_ex2$dem == 0)] <-
  threeC_ex2$age.final65[which(threeC_ex2$dem == 0)]

threeC_ex2$demCR <- threeC_ex2$dem
threeC_ex2$deathCR <- NA
threeC_ex2$deathCR[which(threeC_ex2$dem == 1)] <- 0
threeC_ex2$deathCR[which(threeC_ex2$dem == 0)] <-
  threeC_ex2$death[which(threeC_ex2$dem == 0)]
  
threeC_ex2 <- threeC_ex2 %>% group_by(ID) %>% filter(age.visit65 <= age65_CR)

threeC_ex2 <- threeC_ex2[,c("ID","SBP","age.visit65","num.visit","age0_65",
                            "demCR","deathCR","age65_CR","sex")]
                            

head(threeC_ex2)
ID  SBP age.visit65 num.visit   age0_65 demCR deathCR age65_CR   sex
1  14.2   0.7175222        S0 0.7175222     0       1 2.544216 Woman
1  13.4   0.9576318        S1 0.7175222     0       1 2.544216 Woman
1  12.8   0.9576318        S1 0.7175222     0       1 2.544216 Woman
1  14.2   1.1563997        S2 0.7175222     0       1 2.544216 Woman
1  12.0   1.1563997        S2 0.7175222     0       1 2.544216 Woman
1  11.5   1.1563997        S2 0.7175222     0       1 2.544216 Woman
\end{Code}
Here the individual is considered as dead without dementia at 2.544216.
\subsubsection{LSMM with inter and intra-visit variabilities}
As is the first example, we run the LSMM to initialize the parameters corresponding to the longitudinal sub-model:
\begin{Code}
m2 <- lsmm(formFixed = SBP ~ age.visit65,
           formRandom = ~ age.visit65,
           formGroup = ~ ID,
           timeVar = 'age.visit65',
           data.long = threeC_ex2,
           formVar = "inter-intra",
           random_inter = T,
           random_intra = T,
           formGroupVisit = ~num.visit,
           correlated_re = FALSE,
           S1 = 500,
           S2 = 1000,
           nproc = 4)

summary(m2)
Location-scale linear mixed model fitted by maximum likelihood method 

Statistical Model: 
    Number of subjects: 500 
    Number of observations: 4857 

Iteration process: 
    Convergence criteria satisfied
     Number of iterations:  
          Step 1:  11 
          Step 2:  1 
     Convergence criteria (Step1): parameters = 1.44e-06 
                                 : likelihood = 4.46e-05 
                                 : second derivatives = 1.32e-10 
     Time of computation : 1.427672 mins 

Goodness-of-fit statistics:
    Likelihood:  -8635.61 
    AIC:  17291.22 

Maximum Likelihood Estimates:
Longitudinal model:
      Fixed effects of the location part:
              Coeff     SE    Wald Pvalue
(Intercept) 13.7152 0.1402 97.8547 <0.001
age.visit65  0.3927 0.0874  4.4904 <0.001

     Fixed intercept of the scale part(inter/intra variabilities):
        Coeff     SE     Wald Pvalue
inter  0.2931 0.0275  10.6572 <0.001
intra -0.2216 0.0182 -12.1901 <0.001


     Covariance matrix of the random effects of the mean:
            (Intercept) age.visit65
(Intercept)    3.843600   -1.575881
age.visit65   -1.575881    1.186363

     Covariance matrix of the random effects of the variance:
           inter      intra
inter 0.06036374 0.02102480
intra 0.02102480 0.04600852
\end{Code}

As in the previous example (section \ref{lsmm1}), the \fct{summary} function provides details about the dataset, the convergence process, and the tables of estimates. The \fct{ranef} and \fct{predict} functions computes the predicted random effects \fct{predict}, it also provides the predicted value of the marker.

\subsubsection{LSJM model}
The following lines provide the code to estimate the joint model with competing risks and using the location-scale linear mixed model estimated previously. We assume Weibull functions for both baseline hazard functions. Both risks are adjusted on the sex, the current value of the marker and the between visits variability. The risk of death is also associated on the within visit variability. Data are left truncated by the age of entry in the study \code{age0\_65}:\\ 

\begin{Code}
l2 <- lsjm(Objectlsmm = m2, 
           survival_type = 'CR',
           formSurv_01 = ~ sex, 
           formSurv_02 = ~ sex,
           sharedtype_01 = c("value", "variability inter"),
           sharedtype_02 = c("value", "variability inter",
                                      "variability intra"),
           hazardBase_01 = "Weibull",   
           hazardBase_02 = "Weibull",
           delta1 = ~ demCR, 
           delta2 = ~ deathCR, 
           Time_T0 = ~ age0_65,
           Time_T = ~ age65_CR, 
           nproc = 5,
           S1 = 1000, 
           S2 = 2000)
summary(l2)
Location-scale joint model with competing events 
fitted by maximum likelihood method 

Statistical Model: 
    Number of subjects:  500
    Number of observations:  4857

Iteration process: 
    Convergence criteria satisfied
     Number of iterations:  
          Step 1:  18 
          Step 2:  1 
     Convergence criteria (Step1): parameters = 3.25e-08 
                                 : likelihood = 9.52e-09 
                                 : second derivatives = 1.44e-14 
     Time of computation : 1.526567 hours 

Goodness-of-fit statistics:
    Likelihood:  -9337.714 
    AIC:  18717.428 

Maximum Likelihood Estimates:
Longitudinal model:
      Fixed effects of the location part:
              Coeff     SE    Wald Pvalue
(Intercept) 13.7715 0.1438 95.7902 <0.001
age.visit65  0.3411 0.0918  3.7149 <0.001

     Fixed intercept of the scale part(inter/intra variabilities):
        Coeff     SE     Wald Pvalue
inter  0.3192 0.0286  11.1444 <0.001
intra -0.2185 0.0187 -11.6750 <0.001


     Covariance matrix of the random effects of the mean:
            (Intercept) age.visit65
(Intercept)    3.912766   -1.696949
age.visit65   -1.696949    1.267781

     Covariance matrix of the random effects of the variance:
           inter      intra
inter 0.06732289 0.01799515
intra 0.01799515 0.04611648

Survival models:
    Transition 0-1:
       Regression:
                       Coeff     SE    Wald Pvalue
value 01             -0.1127 0.0915 -1.2313  0.218
variability inter 01  1.7386 0.7074  2.4577  0.014
sexWoman_01          -0.1600 0.2100 -0.7617  0.446

     Baseline:  Weibull 

            Coeff     SE    Wald Pvalue
intercept -4.5011 1.4080 -3.1968  0.001
shape_01   1.6786 0.1091 15.3870 <0.001

    Transition 0-2:
       Regression:
                       Coeff     SE    Wald Pvalue
value 02             -0.0826 0.0512 -1.6143  0.106
variability inter 02  0.9763 0.4540  2.1507  0.032
variability intra 02 -0.6852 0.7087 -0.9669  0.334
sexWoman_02          -0.7785 0.1304 -5.9702 <0.001

     Baseline:  Weibull 

            Coeff     SE    Wald Pvalue
intercept -3.1455 0.8768 -3.5875 <0.001
shape_02   1.9826 0.0674 29.3976 <0.001
\end{Code}

Both risks of dementia and death increase with the inter-visit variability of blood pressure (coeff = 1.7386 and pvalue = 0.014 and coeff = 0.9763 and pvalue = 0.032 respectively) but not with the current value or the intra-visit variability (coeff = -0.6852 and pvalue = 0.334).\\

\begin{figure}[!ht]
    \centering
    \includegraphics[scale = 0.5]{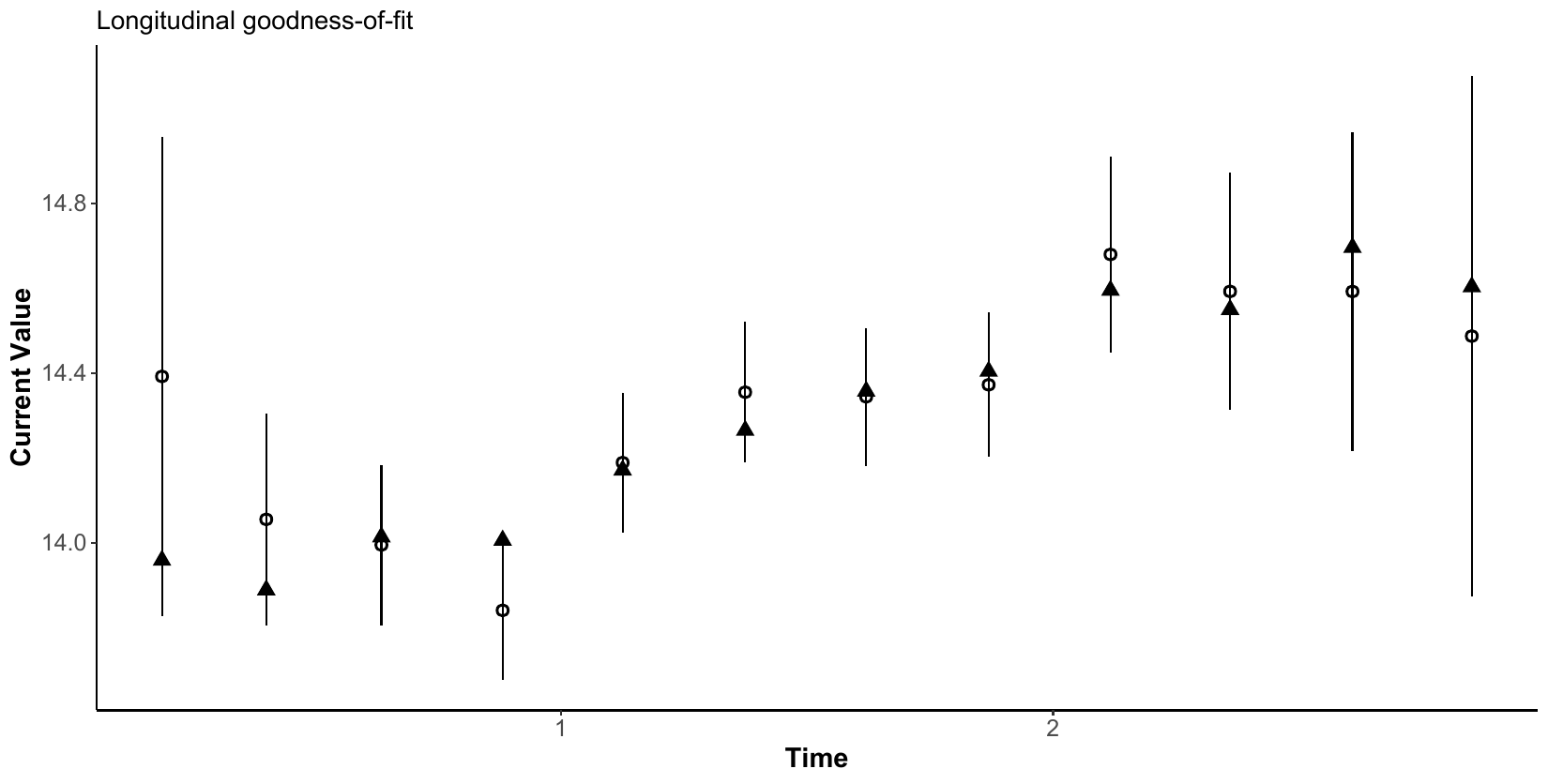}
    \caption{Comparison between predicted values of the marker from the joint model (black triangles) and the observations (mean in white circles with 95\% confidence interval (by 2.5-year age intervals).}
    \label{LongfitEx2}
\end{figure}

\begin{figure}[!ht]
	\centering
	
	\begin{subfigure}{0.48\textwidth}
		\centering
		\includegraphics[width=\textwidth]{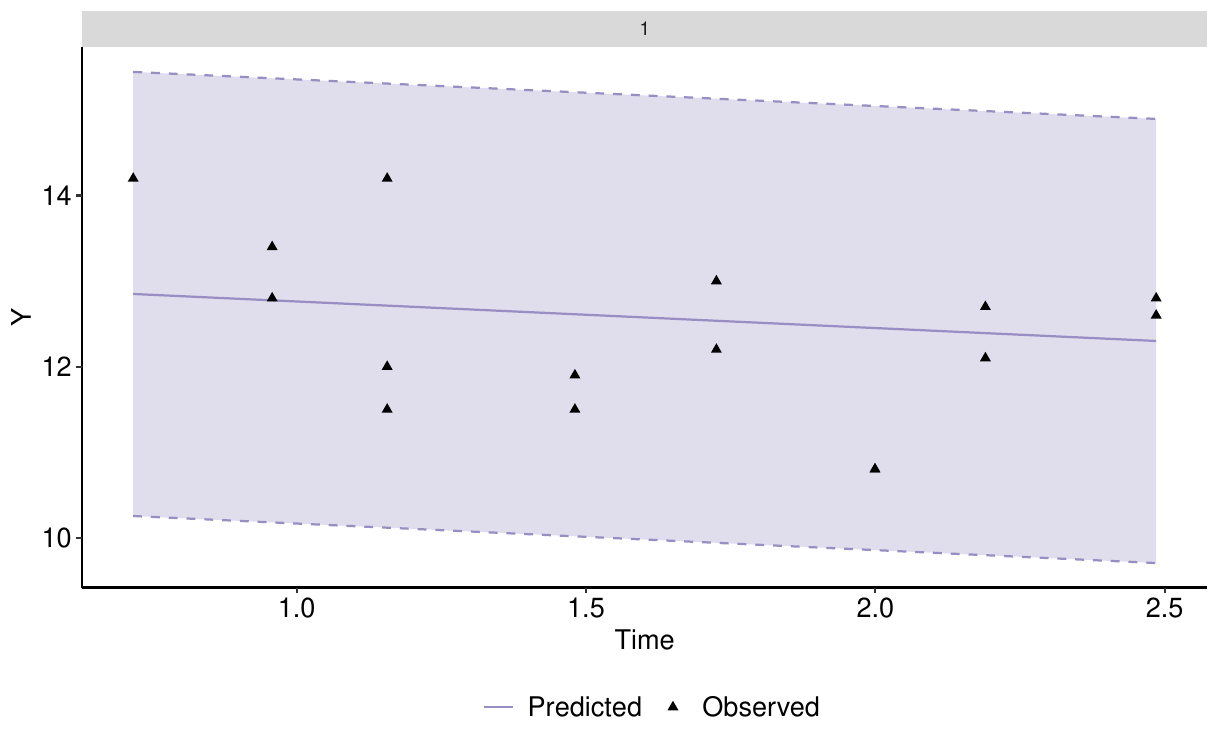}
	\end{subfigure}
	\hfill
	
	\begin{subfigure}{0.48\textwidth}
		\centering
		\includegraphics[width=\textwidth]{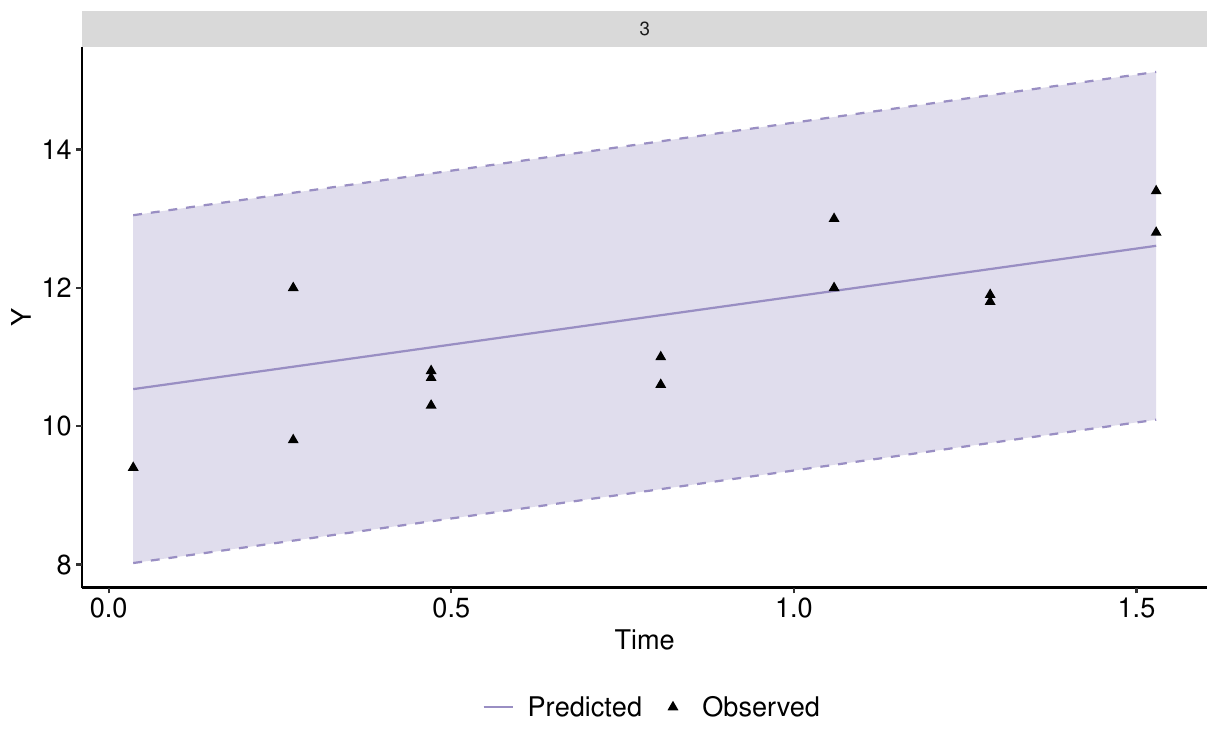}
	\end{subfigure}
	
	\caption{Prediction over time of the individual blood pressure and its prediction interval at 95\% for two subjects. The black triangles are the observed measurements.}
	\label{Ex2.ind}
\end{figure}

\begin{figure}[!ht]
	\centering
	
	\begin{subfigure}{0.48\textwidth}
		\centering
		\includegraphics[width=\textwidth]{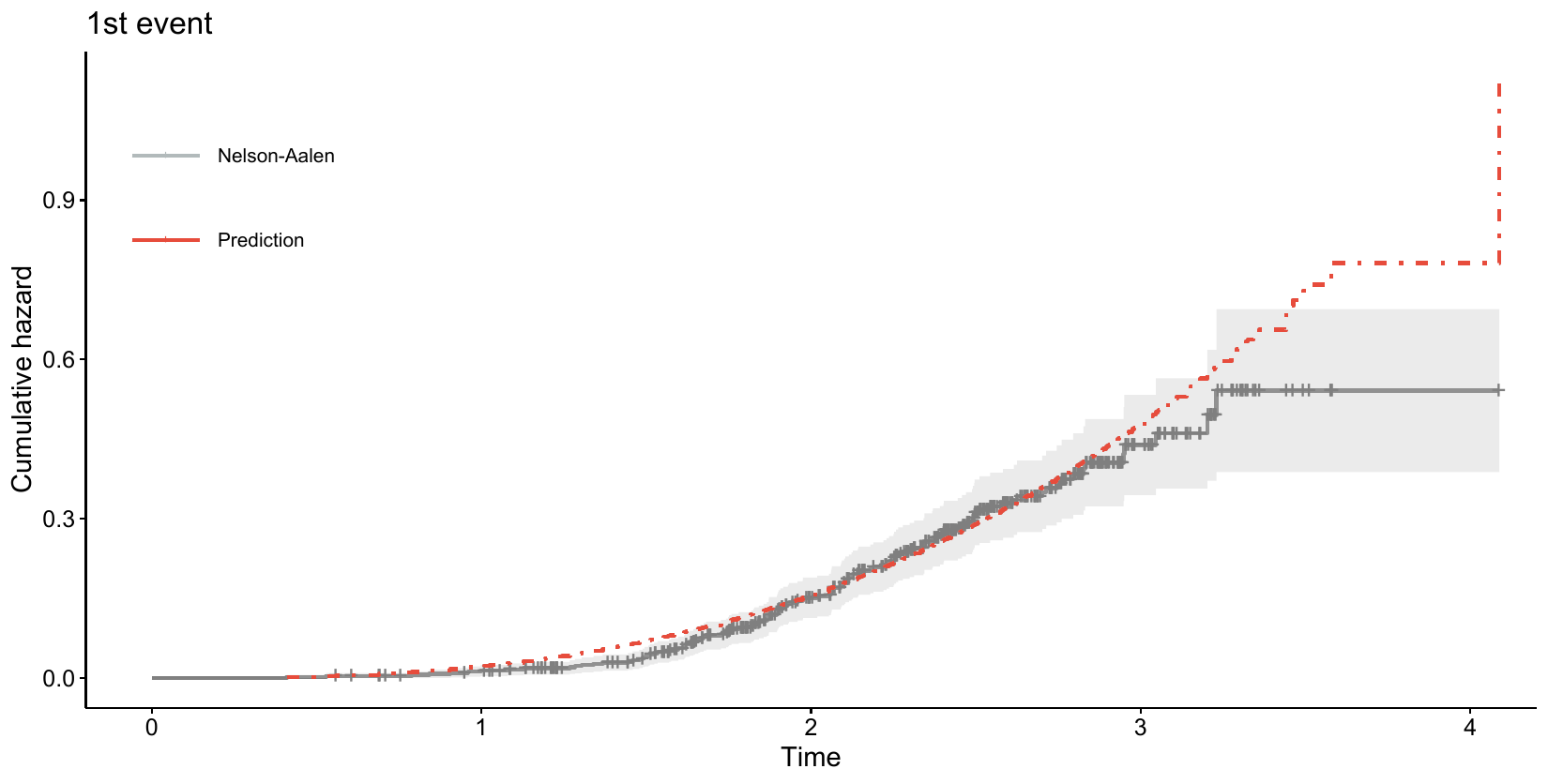}
	\end{subfigure}
	\hfill
	
	\begin{subfigure}{0.48\textwidth}
		\centering
		\includegraphics[width=\textwidth]{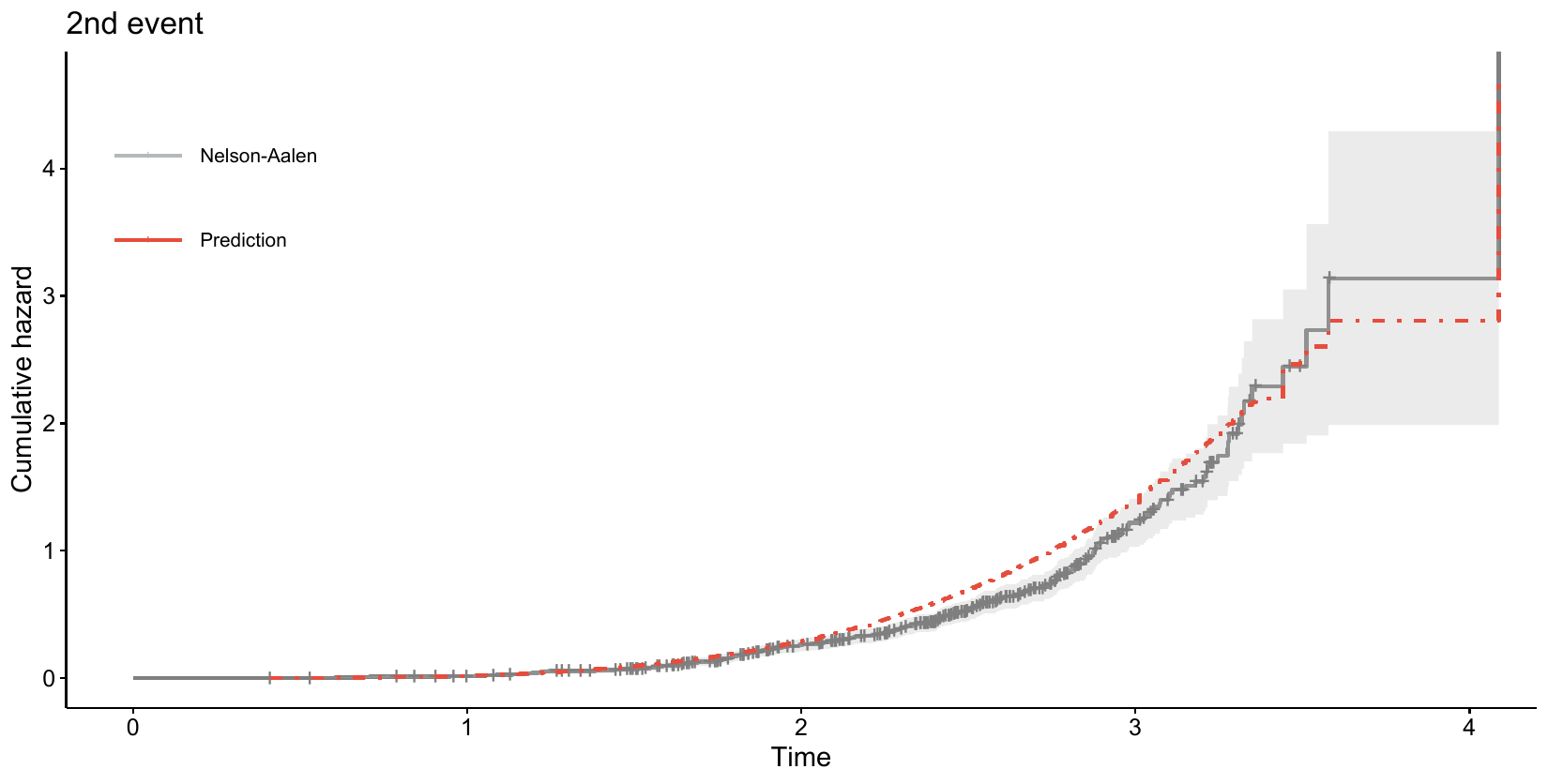}
	\end{subfigure}
	
	\caption{Comparison between predictive cumulative hazard function (in black) and Nelson-Aalen estimator (in red) for the risk of dementia (top) and death (bottom).}
	\label{Ex2.Hazard}
\end{figure}

Then, as in the previous example, \texttt{predict()} function allows to compute the predicted random effects, values of the marker and cumulative functions and \fct{plot} functions displays the Figures \ref{LongfitEx2}, \ref{Ex2.ind}, \ref{Ex2.Hazard} to assess the goodness-of-fit of the model:\\
\begin{Code}
predictl2 <- predict(l2, which = c('RE','Y','Cum'))

plot(l2, which = "long.fit", Objectpredict = predictl2,   
                                 break.times = (seq(65,95,by = 2.5)-65)/10)
plot(l2, which = "traj.ind", Objectpredict = predictl2,  
                                                  ID.ind = c(1,3))
plot(l2, which = "survival.fit", Objectpredict = predictl2)
\end{Code}

\begin{figure}[!ht]
	\centering
	
	\begin{subfigure}{0.48\textwidth}
		\centering
		\includegraphics[width=\textwidth]{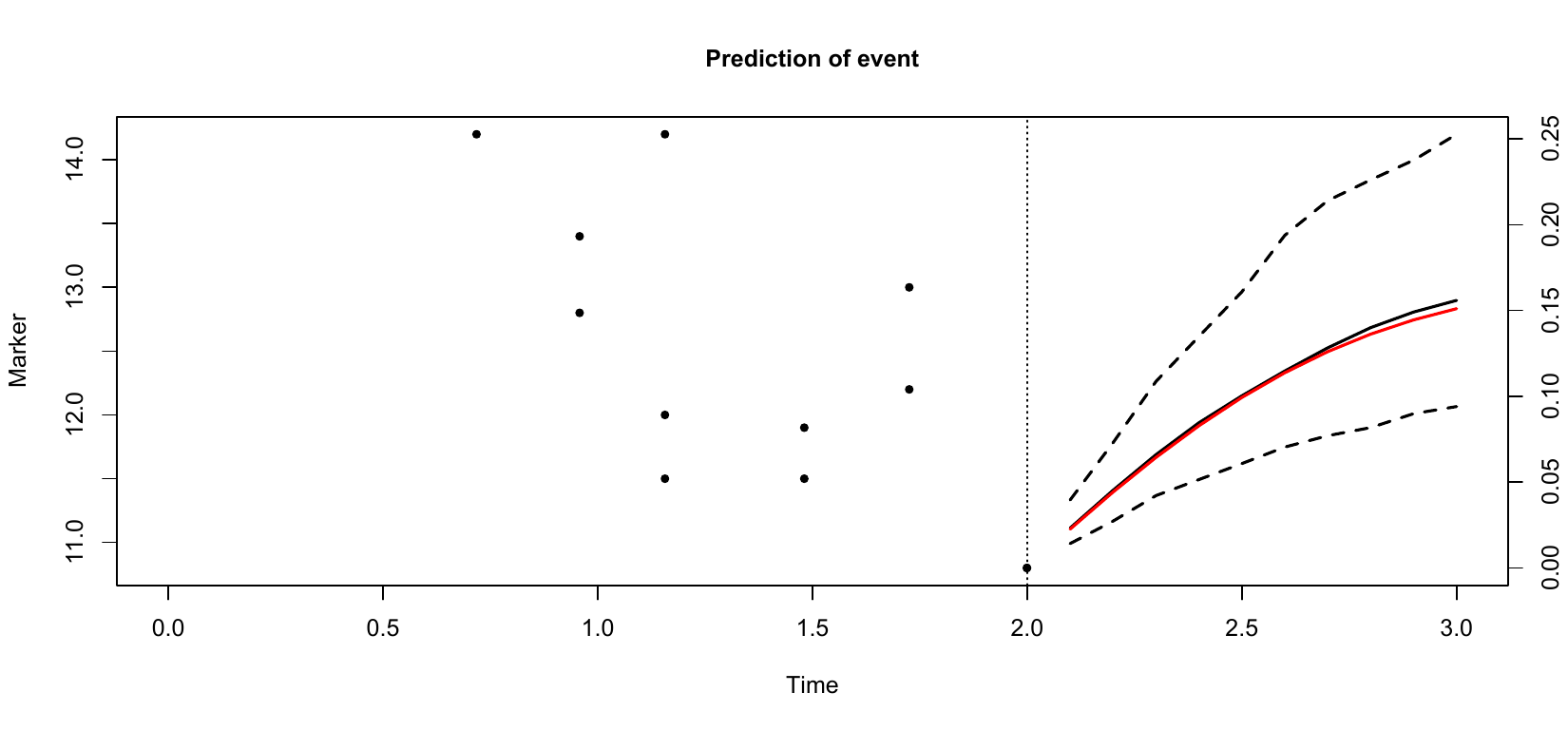}
	\end{subfigure}
	\hfill
	
	\begin{subfigure}{0.48\textwidth}
		\centering
		\includegraphics[width=\textwidth]{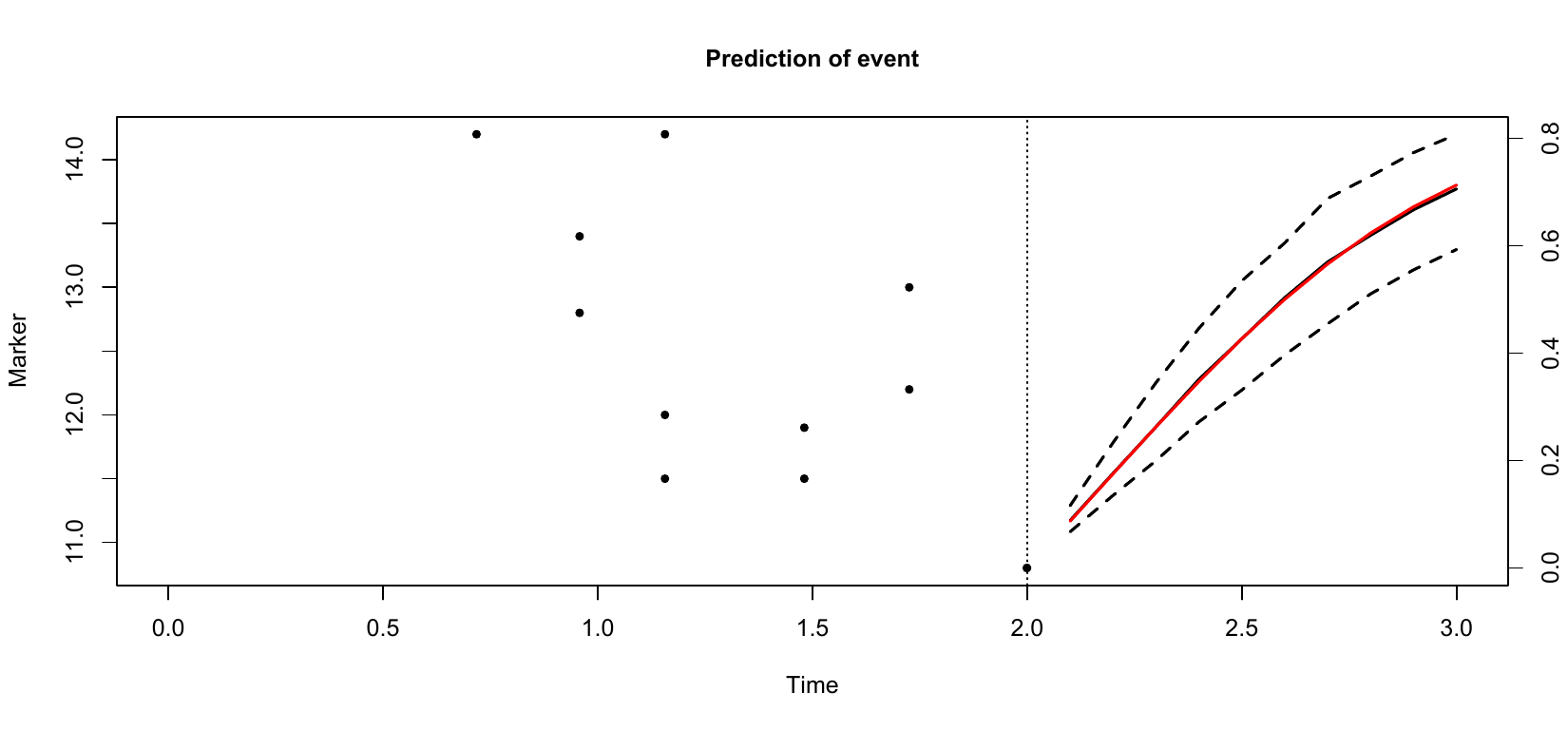}
	\end{subfigure}
	
	\caption{Prediction of the risk of dementia (top) and death (bottom) between 2 and 3 years (with its 95\% confidence interval indicated by dashed lines), for a patient at risk at 2 years.}
	\label{Ex2.Dyn}
\end{figure}

Then, for a competing risk LSJM, dynamic predictions could be computed for both risks:
\begin{Code}
ind2 <- threeC_ex2[which(threeC_ex2$ID == 1),]
dynpDementia <- dynpred(l2, ind2,  s = 2, horizon = seq(2.1,3,0.1),
                                            event = 1, nb.draws = 1000)

dynpDeath <- dynpred(l2, ind2,  s = 2, horizon = seq(2.1,3,0.1),
                                            event = 2, nb.draws = 1000)

\end{Code}
Figures \ref{Ex2.Dyn} display the computed prediction of dementia and death between 2 and 3 years.

Finally, we can also compute the population-average probability of being free of any events at a given time $t$ for a specific profile of covariate:

\begin{Code}
threeC_ex2.id <- threeC_ex2[!duplicated(threeC_ex2$ID),]
# Woman 
indiv_Woman <- threeC_ex2.id[1,]
prob_Woman <- c()
for(t in seq(0.5,3,0.1)){
	prob_Woman <- c(prob_Woman, survmarg(l2, indiv_Woman,t ))}
# Man 
indiv_Man <- threeC_ex2.id[27,]
prob_Man <- c()
for(t in seq(0.5,3,0.1)){
	prob_Man <- c(prob_Man, survmarg(l2, indiv_Man,t ))}
	
cols <- c("#7570B3", "#A6761D")
ltys <- c(1, 2)
age_seq <- seq(0.5, 3, 0.1) * 10 + 65
library(tidyr)
df_plot <- tibble(
	age = age_seq,
	prob_Woman = prob_Woman,
	prob_Man = prob_Man
	) %>%
	pivot_longer(
		cols = -age,      
		names_to = "profile",
		values_to = "prob")
df_plot$profile <- factor(df_plot$profile, levels = c("prob_Woman", "prob_Man"))
	
labels_profiles <- c( "Woman", "Man")
library(ggplot2)
ggplot(df_plot, aes(x = age, y = prob, color = profile, linetype = profile)) +
	geom_line(linewidth = 1.2) +
	scale_color_manual(values = cols, labels = labels_profiles) +
	scale_linetype_manual(values = ltys, labels = labels_profiles) +
	labs(
		x = "Age",
		y = "Probability",
		color = "Subject profiles",
		linetype = "Subject profiles",
		title = "Mean probabilities of being alive and free of dementia 
		across profiles") +
	theme_bw() +
		theme(
			panel.grid = element_blank(),
			text = element_text(size = 14),
			plot.title = element_text(size = 16, face = "bold"),
			legend.title = element_text(face = "bold"))
\end{Code}

\begin{figure}[!ht]
	\centering

	\includegraphics[width=\textwidth]{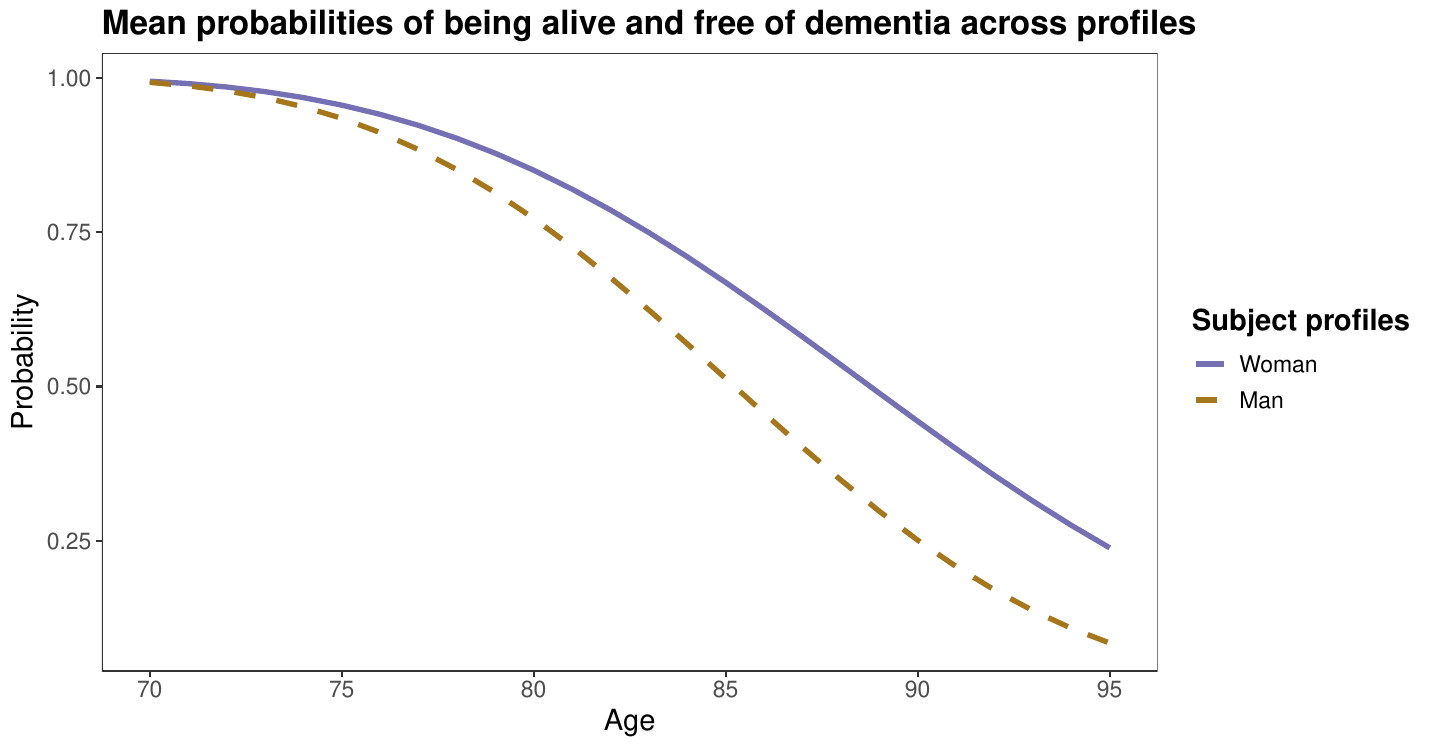}

	\caption{Estimated average probabilities of being alive and without dementia at different
		ages for different profiles of covariates (Women and Men).}
	\label{SurvMargFig}
	
\end{figure}

Figure \ref{SurvMargFig} displays this population-average probability for men and women.

\section{Discussion}
The \pkg{LSJM} package introduces a versatile set of functions designed to estimate new location-scale linear mixed models incorporating various definitions of residual variability, whether it is time-dependent or whether it can be divided into two components, one short-term and one long-term. To our knowledge, no other \proglang{R} package allows all these different fits of the residual variability. It allows also to adjust LSJM with shared random effects combining a LSMM and a complex survival model. The first particularity is the ability to adjust the risk of events based on the individual variability of the marker, in addition to the current value or slope, which is not offered elsewhere. The second is that an illness-death model with interval censoring can be adjusted in this package, something that is not provided by other packages for standard shared random effects joint model. Note that our package can handle the case where residual variability is assumed to be homoscedastic, allowing users to run a classic joint model with or without interval censoring.\\
All of this has been detailed earlier \citep{courcoul_SIM_2025, courcoul2_2024} and illustrated using real \code{threeC} data from the 3C cohort which is included in the package.\\ 
However, the package has some computational limitations, particularly regarding computation time, which can become significantly prolonged when multiple random effects are included. Additionally, convergence issues may arise as the model complexity increases, especially with more intricate random effect specifications. In such cases, increasing the number of QMC draws is often necessary, though this will further extend computation time. Finally, the proposed package allows only one continuous longitudinal marker which is assumed to be Gaussian.\\
Looking ahead, future developments to the \pkg{LSJM} package could include multi-markers models, expanding its capacity to handle more complex datasets. Additionally, extending the package to accommodate location-scale joint models with binary outcomes, as suggested by \cite{elliott_2012}, would further broaden its utility. Then, it would be interesting to extend the estimation to more general family of mixed models either than just the linear one or to consider non linear effects of the current value or the variability on the risk of event. \\
Finally, while the package was initially conceived to analyse blood pressure variability in the context of dementia in a cohort study, as data available in \code{threeC} \citep{courcoul_SIM_2025}, or cardiovascular events in a clinical trial \citep{courcoul2_2024}, its flexible framework makes it applicable to a wide range of other research areas and clinical applications.

\section*{Acknowledgments}
We gratefully acknowledge Catherine Helmer for providing access to the 3C cohort subsample used in this study.

Computer time for this article was provided by the computing facilities MCIA of the Université de Bordeaux and of the Université de Pau et des Pays de l'Adour.

This work was funded by the French National Research Agency (grant ANR‐21‐CE36 for the project “Joint Models for Epidemiology and Clinical research”).

This PhD program was supported within the framework of the PIA3 (Investment for the Future). Project reference: 17‐EURE‐0019.	

\bibliographystyle{plainnat}
\bibliography{Biblio}

\end{document}